\documentclass[reprint,showpacs,preprintnumbers,amsmath,amssymb]{revtex4-1}
\pdfoutput=1
\usepackage[pdftex]{graphicx}
\usepackage{color}
\usepackage{amsmath}
\usepackage{braket}
\usepackage[T1]{fontenc}
\usepackage{url}
\usepackage[english]{babel}
\usepackage{graphicx}
\usepackage{dcolumn}
\usepackage{bm}
\usepackage{subfig}
\usepackage{float}
\usepackage{url}

\newcommand{\be}{\begin{equation}}
\newcommand{\ee}{\end{equation}}
\newcommand{\bea}{\begin{eqnarray}}
\newcommand{\eea}{\end{eqnarray}}
\newcommand{\non}{\nonumber}
\newcommand{\bi}{\begin{itemize}}
\newcommand{\ie}{\item}
\newcommand{\ei}{\end{itemize}}
\newcommand{\mbf}{\mathbf}
\newcommand{\ar}{\arrowvert}

\usepackage{verbatim}

\graphicspath{{figs/}}

\begin{document}

\title{
Chiral symmetry restoration in static-light mesons:
\\ chiral restoration  theorem, the quark running mass $m(k)$ and 
\\ first chiral restoration signals in the lattice QCD spectra.
}

\author{Pedro Bicudo}
\email{bicudo@tecnico.ulisboa.pt}
\affiliation{Lisboa University, CFTP, Dep.\ F\'{\i}sica, Instituto Superior T\'ecnico,  Av.\ Rovisco Pais, 1049-001 Lisboa, Portugal}

\begin{abstract}
Chiral symmetry restoration high in the hadron spectra is expected but it remains to be confirmed both in lattice QCD computations and in experiments. 
Recently, a theorem was derived, relating chiral symmetry restoration high in the hadron spectra to the spontaneous generation of the dynamical quark mass in QCD. We refine the theorem in the case of static-light mesons. 
Utilizing chiral quark model computations and lattice QCD results for the spectrum of mesons composed by a static antiquark and a light quark, we explore chiral symmetry restoration in the  spectrum and the quark running mass $m(k)$. 
\end{abstract}


\maketitle

\section{ Introduction \label{SEC001}}

\subsection{Chiral  symmetry restoration high in the spectrum}

Confinement and spontaneous chiral symmetry (invariance under parity transformation) breaking ($\chi$SB) constitute the main characteristics of QCD at ordinary temperature and density, generating 98 \% of the mass of visible matter in the universe. To understand the mechanisms of $\chi$SB, it is important to study how chiral symmetry can be restored.

Chiral symmetry restoration has been clearly observed in lattice QCD at high temperature, close to the deconfinement crossover \cite{Karsch:2001cy,Aoki:2006we,Kaczmarek:2011zz,Bazavov:2011nk}. In what concerns large density, new chiral phases are expected but they have not yet been observed yet
\cite{Adhikari:2011zf,Kojo:2009ha,Liu:2015lea}. Chiral symmetry restoration is also important for the modifications of the hadron spectrum in nuclear matter
\cite{Bicudo:1993yh,Suzuki:2015est}. Moreover chiral symmetry is the remaining symmetry in the conformal window of technicolour models \cite{Lombardo:2014pda}. Nevertheless, hadron spectroscopy is where, possibly, chiral restoration can be investigated with the highest precision.

In the hadron spectra, chiral symmetry is clearly broken for the groundstates, but we expect chiral symmetry restoration high in the spectrum of hadrons ($\chi$RS) \cite{Cohen:1996zz,Cohen:1996sb,Glozman:1999tk,Cohen:2001gb,Glozman:2007ek,Bicudo:2009cr}, with chiral doublets more and more degenerate,
\be
| M^{P=+} - M^{P=-}| \to 0 \ ,
\label{eq:splitting}
\ee 
as we excite the hadrons.

\begin{figure}[t!]
\includegraphics[width=.85\columnwidth]{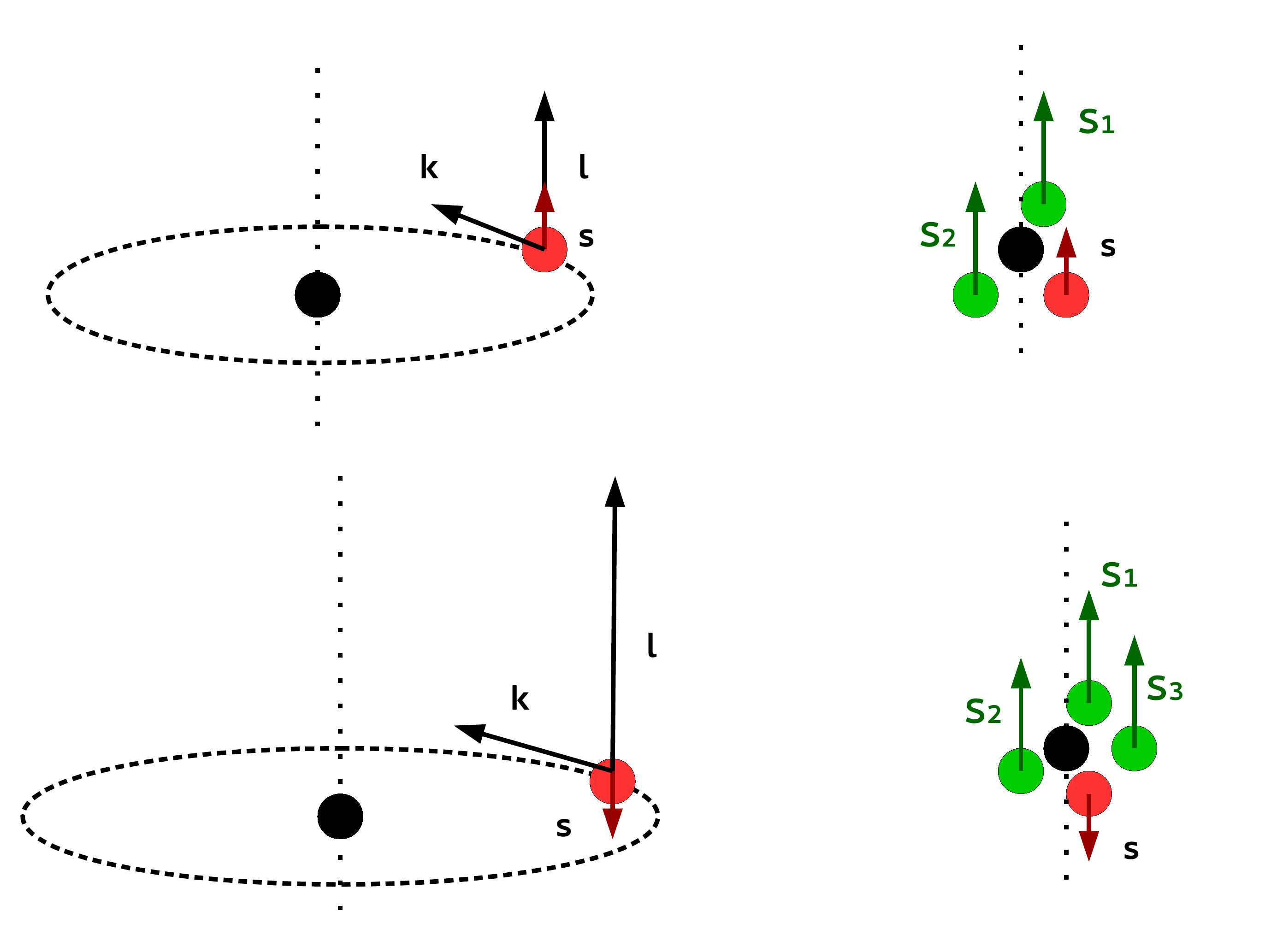}
\caption{\label{fig:01}(Color online). Artist illustration of the antistatic-light system, in a quantum model perspective. In the quark model (left) the angular momentum $\mbf j$  is the sum of the spin $\mbf s$ and orbital angular momentum $\mbf l$ of the light quark and this leads to a large momentum $\mbf p$ for the light quark. If several partons (right) contribute to the angular momentum, it is mostly due to the sum of the parton spins $\sum_ i {\mbf S}_i$, and the light quark has a small momentum. In the top drawings the light quark spin $\mbf s$ is parallel to $\mbf j$ and in the bottom drawings it is anti parallel,}
\end{figure}

Notice $\chi$RS is expected only if the constituent quark degrees of freedom dominate the hadronic system. Other degrees of freedom may be relevant, say the string / gluonic degrees of freedom, or the  valence quark / meson coupled channels degrees of freedom, represented as partons in Fig. \ref{fig:01}, may compete with the radial and angular excitations in the minimal quark degrees of freedom. For instance, the quark spin crisis is an evidence for other relevant degrees of freedom. Thus $\chi$RS, even theoretically, remains an open problem, and deserves to be studied in great detail.

\subsection{The conceptual simplicity of static-light mesons}

Since the  static antiquark - light quark system only has a single light propagator in lattice QCD, it is a conceptually simple hadronic system \cite{Jin:1994ji,Bicudo:1998bz}. It is very convenient to study open problems in hadrons such as
the quark spin crisis 
\cite{Aubert:1985fx},
hybrid excitations 
\cite{Luscher:2002qv}
in the spectrum, suggested for instance by the large degeneracy observed in the meson spectra
\cite{Bugg:2004xu},
 chiral symmetry restoration in the excited spectrum
\cite{Cohen:1996zz,Cohen:1996sb,Glozman:1999tk,Cohen:2001gb,Glozman:2007ek,Bicudo:2009cr}.

To search for a signal of $\chi$RS we thus address chiral symmetry restoration in static-light systems. Static-light systems are ideal mesons where, say the antiquark, is so heavy it freezes, and only the light quark is dynamical. Lattice QCD provides the framework study the ideal static-light mesons.

We consider the system of a static, extremely heavy with mass $m_\text{static}$ and equivalent to spinless, antiquark in the $\bar 3$ representation, and of a standard light quark. In lattice QCD, the only dynamical degrees of freedom in this system are the ones of the light quark and gluons, and the spectrum only depends on the quantum numbers of the light degrees of freedom $j, l, n, s$, as illustrated in Fig. \ref{fig:01}.

In the scenario where all angular momentum and dynamical energy reside in the light quark (left of Fig. \ref{fig:01}) we expect a Regge-like behaviour for the meson mass $M$, with $M -m_\text{static} \propto \sqrt j$. In case the angular momentum and dynamical energy mostly reside in several partons (right of Fig. \ref{fig:01}), we expect and additive behaviour for both the energy and angular momentum, with  $M-m_\text{static} \propto j$. Thus in the large angular momentum limit, we expect the component of the Fock space with a single light quark and no partons to dominate the system, and this is the scenario we will adopt from now on. This leads to  $\chi$RS, thus if there is no observed $\chi$RS the pure quark-antiquark modelling of static-light systems should be abandoned.

In case one would like to compare with an actual heavy-like meson 
\cite{Agashe:2014kda}, 
one would only need to add the heavy antiquark spin, resulting in a $\pm 1/2$ shift in the total angular momentum $\mbf J=  \mbf j + \mbf s_\text{\,heavy}$.

\subsection{The mechanism of $\chi$RS, and  $\chi$RS as a probe for the running quark mass }

The only term breaking chiral symmetry in the QCD lagrangian is the mass term $\bar \psi \, m_\text{current} \, \psi$, it is not invariant under flavour $SU_A(N_f)$ chiral rotations $ \psi \to \exp{ i \theta \gamma_5 \lambda^a}$.  $U_A(1)$ is also broken due to the chiral anomaly. The current quark quark masses are UV renormalized in the standard model, and the result is that the light flavours $u, \ d$ have very light current masses when compared with the finite scale of QCD (even the $s$ quark can be approximated as a light quark). The current quark mass is of the MeV order, i e for the quark up $m_\text{current}=2.3^{+0.7}_{-0.5}$ MeV and for the quark down $m_\text{current}=4.8^{+0.7}_{-0.3}$ MeV  \cite{Agashe:2014kda}. 

\begin{figure}[t!]
\includegraphics[width=.85\columnwidth]{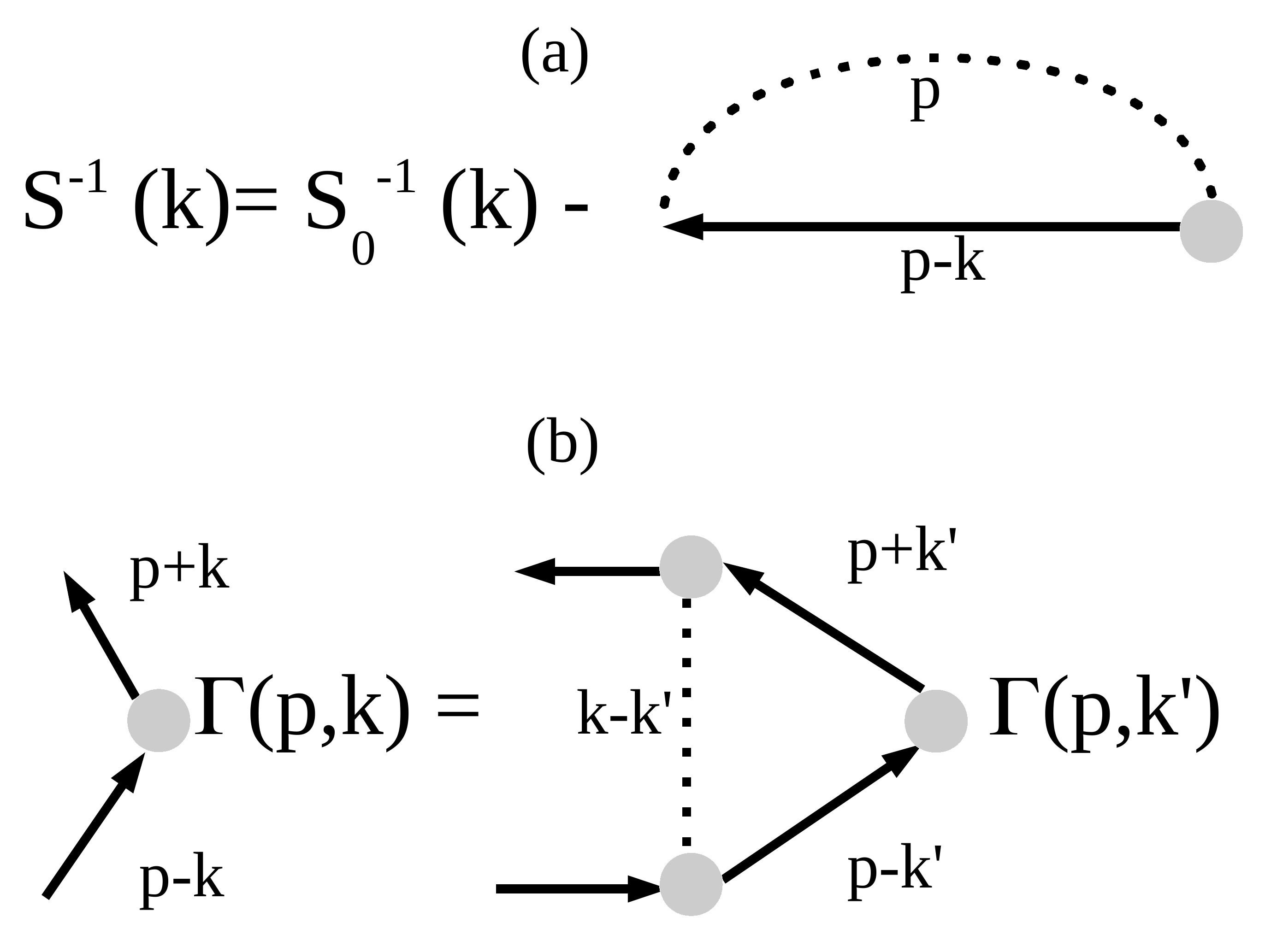}
\caption{\label{fig:DysonSchwinger} Diagramatic version of the dyson-Schwinger equations, (a) the mass gap equation and (b) bound state equation. The solution of (a) is the running quark mass $m(k)$ present in the full quark propagator $S$, whereas the free quark propagator $S_0$ only depends on the current quark mass $m_\text{current}$. In the diagrams, the solid line is the quark propagator, the dotted line represents the quark-antiquark interaction and the disks represent the effective vertices of the interaction, and the Bethe Salpeter vertices of the boundstates.}
\end{figure}

Besides, spontaneous $\chi$SB also occurs.  In lattice QCD, $\chi$SB is directly measured with the non-invariant observable $\langle \bar \psi \psi \rangle$, and the lattice QCD evidence is that it occurs in coincidence with confinement
\cite{Karsch:2001cy,Aoki:2006we,Kaczmarek:2011zz,Bazavov:2011nk}. In the continuum perspective, $\chi$SB occurs via the solution of the non-linear Dyson-Schwinger equation for the quark propagator of Fig. \ref{fig:DysonSchwinger} (a). The propagator includes a dynamically generated quark mass $m(k)$, a function of the quark momentum $k$, also known as constituent quark mass. Both perspectives are related to confinement, and to the infrared scale of QCD, say the scale of the string tension $\sigma$. The Dyson-Schwinger approach is similar to the renormalization program, however the quark and antiquark interaction leading to $m(k)$ is the confining one (not the UV interaction) and the equations resulting from the diagrams are non-linear (already at one loop). This results in multiple solutions for $m(k)$. Another approach, in the Fock space with a Bogoliubov-Valatin transformation, equivalent to Dyson-Schwinger, shows the solution with the largest $m(k)$ is the one corresponding to the stable, physical vacuum \cite{Bicudo:1989sh}, and the other solutions include the perturbative vacuum obtained with UV calculations only. The constituent quarks mass, in phenomenological quark models \cite{Godfrey:1985xj}, is typically two orders of magnitude larger than the current quark mass, $m(0) \sim 300$ to $400$ MeV.

QCD, including the spontaneous $\chi$SB generation of the quark mass $m(k)$, accounts for 98 \% of the visible mass in the universe. This mass also breaks $\chi$SB in the spectra of hadrons. In particular, in the Bethe-Salpeter, or Dirac equations for QCD boundstates of Fig. \ref{fig:DysonSchwinger} (b), when the vertices are directly computed from QCD and are chiral invariant, $m(k)$ is the only term leading to the mass splitting in chiral multiplets. Clearly, chiral symmetry is broken in the light hadron spectrum, where the parity doublet splitting in Eq. (\ref{eq:splitting}) is particularly large for light hadrons. For instance the splitting between the pseudoscalar $\pi$ and scalar $a_0$ mesons, or between the vector $\rho$ and axial vector $a_1$ or $b_1$ mesons is also of the order of $\sim 500$ to 800 MeV $\sim 2 \, m(0)$.

For simplicity let us illustrate $\chi$RS with the quark propagator (already dressed with the Dyson-Schwinger equations). Let us consider the light quark full propagator can be parametrized as,
\be
S(k) = { i  \ Z(k) \over \not k -m(k) + i \, \epsilon}\ .
\label{propag}
\ee
The only term breaking chiral symmetry is the dynamically generated quark mass $m(k)$. The quark propagator can be computed in  Lattice QCD, when gauge fixing is implemented. In the Dyson-Schwinger formalism, the mass $m(k)$ is the solution of the self energy, or mass gap, equation illustrated in Fig. \ref{fig:DysonSchwinger} (a) . The Dyson-Schwinger formalism, even when truncated at the ladder approximation,  is consistent with the low energy chiral theorems \cite{Bicudo:2003fp}. 
In light hadrons the average momentum of a constituent quark is of the same order of its constituent quark mass. However,  in systems where
\be
 	\left< {m(k) \over k}  \right> \to 0 \ ,
 \label{eq:massmomentum}
 \ee
in particular in excited hadrons, where we expect the average momentum of the quark $ \langle k \rangle$ to increase with the excitation level, we expect the constituent mass $m(k)$ to be negligible, leading to $\chi$RS.

Lattice QCD computations \cite{Lang:2011vw,Lang:2011ai,Schrock:2011hq,Burgio:2012ph,Glozman:2012fj,Schrock:2013xf}
indicate the quark mass tends to vanish with an infrared momentum cutoff, interpolating between the constituent quark mass $m$ at small momentum and the current quark mass at large momentum,
 \be
 \lim_{k \to \infty} m(k) \to m_\text{current} \ ,
 \ee
 where $m_\text{current} << m $. 
This is promising for $\chi$RS, since a possible decrease of the quark mass would also go in the direction of Eq. (\ref{eq:massmomentum}). Possibly a finite number of hadronic excitations could be sufficient to already observe $\chi$RS. Indeed, the quark propagators, with a dynamical generated quark mass, computed in any known approach \cite{Szczepaniak:1996gb}, i e in gauge fixed lattice QCD, gauge fixed Dyson-Schwinger equations, or in chiral invariant quark models, all predict a quark mass decreasing with momentum. 
Thus, according to a recent theorem \cite{Bicudo:2009cr} an observation of $\chi$RS would be welcome to discriminate which running quark mass is quantitatively correct. It should provide a gauge invariant definition of the running quark mass, interpolating between the constituent and current quark masses.

\begin{figure}[t!]
\vspace*{-15pt}
\centerline{\includegraphics[height=190pt]{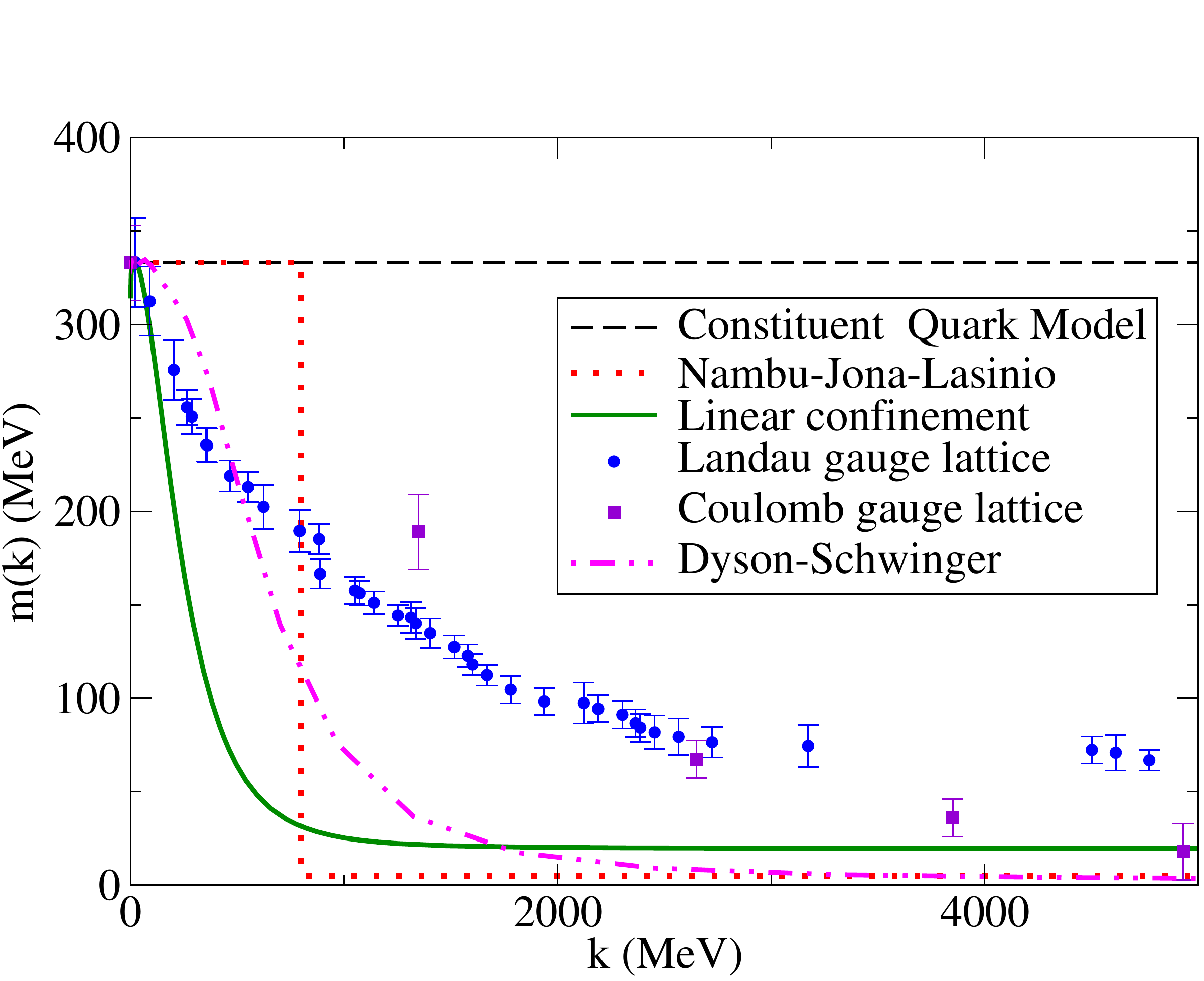}}
\caption{IR enhancement of the light quark mass due to spontaneous
  $\chi$SB \cite{Bicudo:2009cr}. Shown are quark masses in the main approaches to QCD, all
  multiplied by an arbitrary factor to match them at quark momentum norm $k \to
  0$. We thank the authors of Ref. \cite{Bicudo:2009cr} for this figure. 
\label{fig:massgeneration}}
\vspace*{-15pt}
\end{figure}

However, not only $\chi$RS is not yet settled, moreover the different theoretical approaches computing the constituent quark mass produce very different functions $m(k)$, as shown in  Fig. \ref{fig:massgeneration}. The discrepancies occur because the mass gap equation is extremely non-linear, and the solution $m(k)$ depends strongly on the details of confinement.
For a definitive understanding of $\chi$RS and of the running quark mass, experiment or lattice QCD are necessary,

In lattice QCD, an indication of chiral symmetry restoration has already been shown when an infrared cutoff is imposed. 
When the quark moment is forced to increase arbitrarily, approaching the limit $k \to \infty$, lattice QCD computations 
\cite{Lang:2011vw,Lang:2011ai,Schrock:2011hq,Burgio:2012ph,Glozman:2012fj,Schrock:2013xf}
show that hadrons group in degenerate chiral multiplets as in Eq. (\ref{eq:splitting}). However, no direct evidence for $\chi$RS has been reported previously in lattice QCD, although it is possible to compute the static-light spectrum in lattice QCD, with the correlators shown in Fig. \ref{fig:04}.

In what concerns experiment, $\chi$RS is not settled yet. In general it is difficult to observe highly excited hadrons. There is evidence from the former Cristal Barrel collaboration at CERN \cite{Aker:1992ny,Anisovich:2000ut,Anisovich:2011sva,Anisovich:2001pn,Anisovich:2002su} of a symmetry  in the excited meson spectra \cite{Bugg:2004xu,Bicudo:2007wt,Denissenya:2014poa,Cohen:2015ekf}, but it has not been confirmed yet by other experiments.
 Moreover in mesons, and even more so in hadrons, the spectra have many states with similar quantum numbers, and parity partners are not straightforward to identify.

\subsection{summary}

We extend the theorem predicted in Ref.  \cite{Bicudo:2009cr} to antistatic-light mesons. 
We expect the conceptual simplicity of  antistatic-light mesons, together with a possibly favorable running quark mass  $m(k)$ may produce  a signal of $\chi$RS. 
We are thus interested in studying possible trends in the mass splitting $\ar M^+\! -\! M^- \ar$ as a function of increasing total angular momentum of the light quark $ j$.  

In Section \ref{SEC002} we specialize the QCD inspired theorem \cite{Bicudo:2009cr}, deriving it for the case of $\chi$RS in static-light mesons. We show how, from the spectrum of static-light mesons, a power law relation for the running quark mass $m(k)$ is extracted. 

In Section \ref{SEC003} we review chiral symmetric quark models and apply them to static-light mesons.
We test our theorem in the particular case of the quadratic and linear confining quarks models.  

In Section \ref{SEC004} we review the angular excitations of the antistatic-light system, in lattice QCD
\cite{Baron:2009wt,Jansen:2008si,Michael:2010aa,Wagner:2011fs,Foley:2007ui} .
We discuss the lattice QCD spectrum for static-light mesons computed with dynamical quarks and compare them with previous lattice QCD  computations with quenched light quarks. We compare the lattice QCD results with our theorem, and extract the trend for the running mass $m(k)$ of the light quarks. 

Most of our work builds on references \cite{Bicudo:2009cr,Bicudo:1998bz,Foley:2007ui,Jansen:2008si,Michael:2010aa}. We finally conclude in Section \ref{SEC005}.

\section{A QCD inspired theorem for $\chi$RS  in the direction of large angular momentum $j$ of the light quark in the static-light meson. \label{SEC002}}

We now aim at a model independent theorem for the $\chi$RS in static-light systems at large  large angular momentum $j$.
We first decompose the Dirac quark spinors with $ 4 \times 4 $  indices, in quark and antiquark operators, each with $2 \times 2 $  spin indices. This is enables us to decompose the Feynman-Dirac quark propagator of Eq. (\ref{propag}),
\begin{eqnarray}
\label{DiracWeyl}
{\cal S}_{Dirac}(k_0,\vec{k})
&=&u(\mbf k)
{\cal S}_{q}(k_0,\vec{k})
u^{\dagger}(\mbf k) \beta
\nonumber \\
&&- v^{\dagger}(\mbf k)
{\cal S}_{\bar q}(-k_0,-\vec{\mbf k})
v(k) \beta  \ ,
\end{eqnarray}
where the quark and antiquark Weyl-Goldstone propagators are.
\begin{equation}
{\cal S}_{q}(k_0,\vec{k})={\cal S}_{\bar q}(k_0,\vec{k})=
{i \over k_0-E(k) +i\epsilon} \delta_{s \, s'} \ ,
\end{equation}
and the quark and antiquark spinors are decomposed in, 
\begin{eqnarray}
u_s({\bf k})&=&\left[
\sqrt{ 1+S \over 2} + \sqrt{1-S \over 2} \widehat k \cdot \vec \alpha
\right]u_s(0)  
\nonumber \\
v_s({\bf k})&=&\left[
\sqrt{ 1+S \over 2} - \sqrt{1-S \over 2} \widehat k \cdot \vec \alpha
\right]v_s(0)   \ ,
\end{eqnarray}
where $s=\pm 1/2 $ is the quark or antiquark spin subindex. 
This is convenient to derive, from the Bethe-Salpeter boundstate equation in Minkowski momentum space, 
a Hamiltonian formalism, where the quantum numbers for the light quark are similar to the ones of the quark model, i. e. 
its spin $\mbf s$, radial angular momentum $\mbf l$ and total angular momentum $\mbf j$.
More detail on this decomposition is found in Ref. \cite{Bicudo:1998mc} and Refs. therein.

 To derive our theorem, we now follow closely Ref. \cite{Bicudo:2009cr}, denoting 
$ H^{QCD\ '}_{\chi}$ as a term in the Hamiltonian of first order in the $ \langle {m(k) \over k} \rangle$ expansion. We also utilize units of $\hbar = c = 1$.
We get,  
\be
\label{eq:paritysplitofm}
\left| M^+ - M^- \right|
\to 
 \langle { m( k )\over k } H^{QCD\ '}_{\chi} \rangle \ .
 \ee 
In what concerns the term  $H^{QCD\ '}_{\chi} $, for general hadrons, say mesons or baryons, it may take different forms.
Nevertheless in the static-light system, since there is only one unfrozen quark spin and only one unfrozen quark momentum,
we are able to determine its behaviour. The only spin-dependent potential, separating the $P=+1$ from the $P=-1$ cases, and actually the only operator able to interpolate between the massive angular momentum barrier and the $j$ dependent only potential of Eq. (\ref{resto}), is the spin-orbit potential ${\bf s}\cdot{\bf l} $. Moreover assuming a linear confinement $\sigma r$, which dominates the boundstate wavefunctions at large momentum  of Fig. \ref{fig:DysonSchwinger} (b), a dimensional analysis leas to a momentum dependence $\sigma / k$. Thus we arrive at, 
\be
 H^{QCD\ '}_{\chi}  \to  c \,  {\bf s}\cdot{\bf l}  { \sigma  \over k} \ .
 \label{eq:Hqcd}
\ee
where $c$ is a simple constant we do not attempt to determine here.

The spin-orbit term ${\bf s}\cdot{\bf l} = { j(j+1) -l(l+1) -s(s+1) \over 2}$. Note the light quark parity is $p=(-1)^l$, whereas the static antiquark has an intrinsic parity  $-1$, $P=- p$. In our case $l=j \pm {1 \over 2}$, thus the spin-orbit term is the one breaking chiral symmetry in the spectrum, 
\be
\left| \langle {\bf s}\cdot{\bf l}  \rangle^+ -  \langle {\bf s}\cdot{\bf l}  \rangle^- \right| \to j \ ,
\ee
and thus we get,
\be
\left|  M^+- M^- \right| \to c \, \sigma \, j \langle {  m(k) \over k ^2 }  \rangle \  .  
\label{eq:sldone}
\ee

\begin{figure}[t!]
\includegraphics[width=.85\columnwidth]{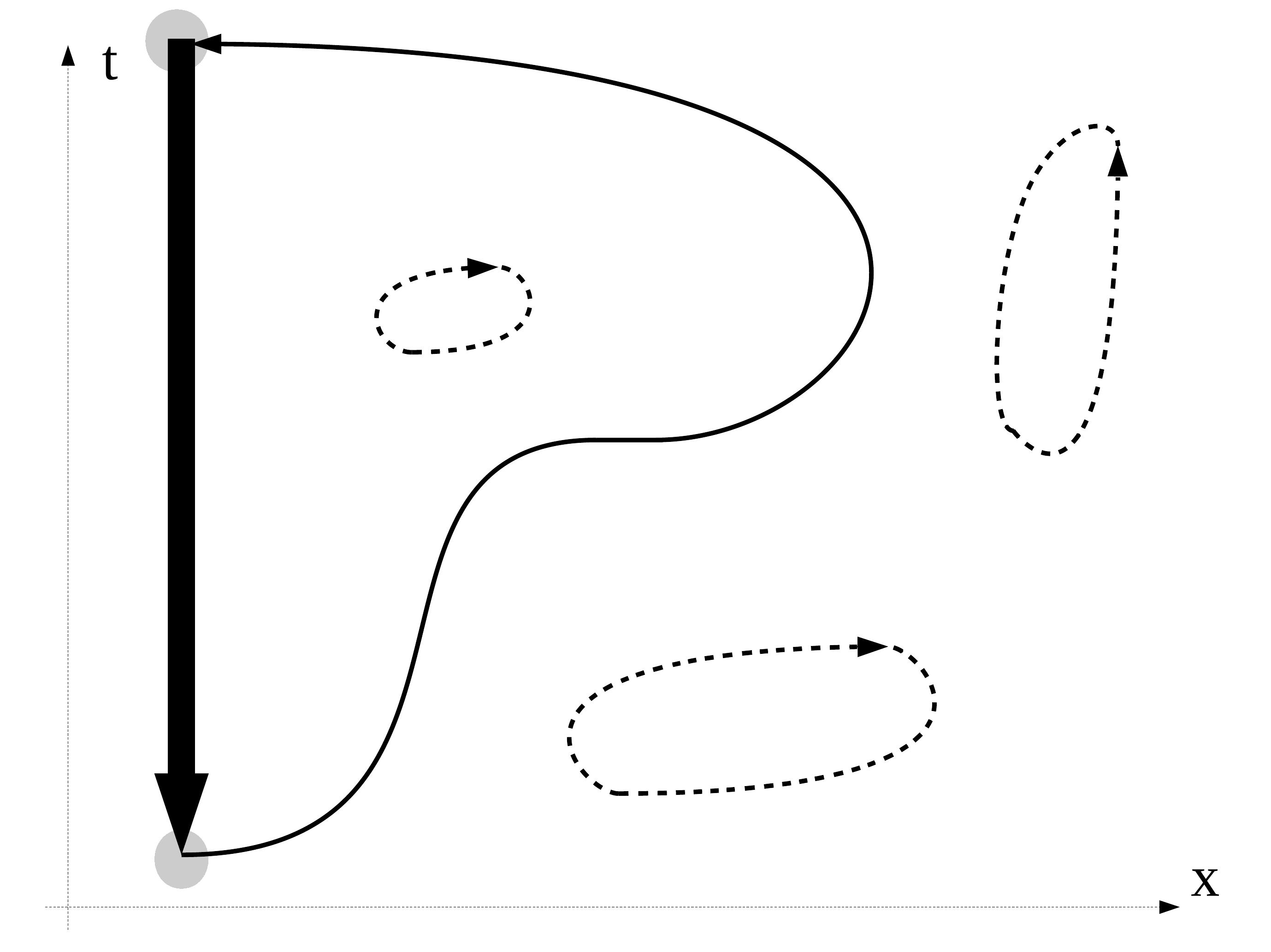}
\caption{\label{fig:04}. Illustration of the antistatic-light system, in a lattice QCD perspective, where the anti quark static source is a Wilson temporal line (thick straight arrow) and the light valence quark propagator (thin arrowed lines) is used as well. Moreover, possible sea quark loops are present as well (thin dashed loops), any possible gluonic excitations are also included (not represented in the figure), and lattice QCD includes all possible contributions to the angular momentum $\mbf j$ illustrated in Fig. \ref{fig:01}.}
\end{figure}

To proceed with our demonstration, we change variable from the momentum $k$ to the angular momentum of the light quark $j$. 
We assume  the $| M^+ - M^-| $ mass difference is decreasing. In the limit of large momenta , the kinetic energy of the light quark tends to $k$ and the spin dependent terms are smaller than the kinetic energy and the potential (including the centrifugal barrier).  The Hamiltonian is essentially equivalent to $H \to k + \sigma r$ where $\sigma$ is the string tension of the linear confinement. 
Applying the relativistic  virial theorem \cite{Lucha:1990mm}, we find the kinetic energy and the potential energy are both  proportional to the total energy of the light quarks,
\be
\langle k \rangle \sim \langle \sigma r \rangle \to  {1 \over 2}  M^\pm \ .
\label{eq:averagemomentum}
\ee
Moreover, there is evidence light hadrons are in linear Regge trajectories, consistent with the linear confining potential
\cite{Agashe:2014kda}, for instance in the leading trajectory of light mesons with $J^{PC}$ of $1^{--},\,2^{++},\,3^{--},\,4^{++}$. 
In the case of static-light mesons, the trajectories are not studied yet.
Nevertheless we expect the light quark energy $M^{\pm} -m_\text{static}$ to be in a Regge trajectory,
since the dynamics of the bound state equation for one light quark is similar, in the Salpeter equation,
to the dynamics of the bound state equation for a meson. 
In particular, both have a linear confining potential.
The two approximately degenerate masses $M^+$ and $M^-$ are
in the same leading linear Regge trajectory, phenomenologically fixing
their $j$-dependence to 
\be 
\label{eq:trajectory} 
j = \alpha_0 + \alpha
\left( {M^\pm} -m_\text{static} \right)^2
\mathop {\longrightarrow }\limits_{j \to \infty }
\alpha \left( {M^\pm} -m_\text{static} \right)^2 \ ,
\ee 
where $\alpha$ is the respective Regge slope.

Combining Eqs. (\ref{eq:sldone}) ,  (\ref{eq:averagemomentum}), (\ref{eq:trajectory}) and assuming for large angular momenta the momentum distribution has a small variance, consistent with a single peak,
we finally find
\be
\left| M^+ - M^- \right| \to  c {4 \sigma \over \alpha } m \left( { 1 \over 2 \sqrt \alpha} \sqrt j \right)  \ .
\label{eq:theorem}
\ee
Thus plotting the parity mass splitting $\left| M^+ - M^- \right|$ as a function of $\sqrt j$ should reproduce the momentum dependence of the running quark dynamical mass as a function of momentum. Notice, since $\left| M^+ - M^- \right|$ is an observable of lattice QCD, this result is gauge invariant.

\begin{table*}[t!]
\caption{Meson mass $M$ in  units of the potential parameter $K_0$ as obtained 
\cite{Bicudo:1998bz} from Eq. (\ref{eq:bs}) in the chiral limit of current quark mass $m_\text{current} \to 0$. $j^ p, \ l$ are respectively the total angular momentum, the parity, and the orbital angular momentum of the light quark. $J^P$ is the corresponding angular momentum and parity of the static-light meson, including the spin $1 \over 2$ of the heavy quark. We compute the spectrum up to $J^p= 39/2	\,^\pm$ but only show the table up to  $13/2 	\,^\pm$ because the parity doublets with higher angular momenta are essentially degenerate.}
\label{tab:quadratic}
\begin{ruledtabular}
\begin{tabular}{|c|c|c|r|r|r|r|r|r|r|r|r|r|}
 $j^p$    &   $l$    & $J^P$ & $n=0$   & $n=1$   & $n=2$ & $n=3$   & $n=4$   & $n=5$ & $n=6$   & $n=7$   & $n=8$ & $n=9$    \\
\hline
 ${1 / 2 }	\,^+$       & 0        &  $0^- \ , \ 1^-$  & 2.599 & 4.913 & 6.767 & 8.390 & 9.868 & 11.242 & 12.535 & 13.764 & 14.940 & 16.071  \\
 ${1 / 2 }	\,^-$      & 1        &  $0^+ \ , \ 1^+$ & 3.272 & 5.460 & 7.254 & 8.841 & 10.292 & 11.646 & 12.922 & 14.139 & 15.303 & 16.425 \\
 ${3 / 2 }	\,^-$      & 1        &  $1^- \ , \ 2^-$ & 4.545 & 6.383 & 8.004 & 9.484 & 10.860 & 12.158 & 13.391 & 14.571 & 15.707 & 16.803  \\
 ${3 / 2 }	\,^+$       & 2        &  $1^+ \ , \ 2^+$ & 4.616 & 6.487 & 8.127 & 9.618 & 11.003 & 12.307 & 13.544 & 14.728 & 15.866 & 16.964  \\
 ${5 / 2 }	\,^+$       & 2        &  $2^- \ , \ 3^-$ &  5.732 & 7.415 & 8.938 & 10.347 & 11.670 & 12.925 & 14.122 & 15.273 & 16.382 & 17.456  \\
 ${5 / 2 }	\,^-$      & 3        &  $2^+ \ , \ 3^+$  & 5.744 & 7.437 & 8.969 & 10.385 & 11.714 & 12.974 & 14.176 & 15.330 & 16.443 & 17.520  \\
 ${7 / 2 }	\,^-$      & 3        &  $3^- \ , \ 4^-$  & 6.749 & 8.320 & 9.766 & 11.119 & 12.397 & 13.616 & 14.784 & 15.909 & 16.997 & 18.052  \\
 ${7 / 2 }	\,^+$       & 4        &  $3^+ \ , \ 4^+$  & 6.752 & 8.325 & 9.775 & 11.130 & 12.412 & 13.633 & 14.804 & 15.931 & 17.021 & 18.077   \\
 ${9 / 2 }	\,^+$       & 4        &  $4^- \ , \ 5^-$  & 7.679 & 9.162 & 10.545 & 11.848 & 13.088 & 14.274 & 15.415 & 16.517 & 17.585 & 18.621   \\
 ${9 / 2 }	\,^-$      & 5        &  $4^+ \ , \ 5^+$  & 7.680 & 9.164 & 10.547 & 11.852 & 13.093 & 14.280 & 15.422 & 16.525 & 17.594 & 18.632   \\
 ${11 / 2 }	\,^-$     & 5        &  $5^- \ , \ 6^-$  & 8.550 & 9.962 & 11.291 & 12.551 & 13.756 & 14.913 & 16.029 & 17.109 & 18.157 & 19.177  \\
 ${11 / 2 }	\,^+$      & 6        &  $5^+ \ , \ 6^+$  & 8.550 & 9.963 & 11.291 & 12.552 & 13.757 & 14.915 & 16.031 & 17.112 & 18.161 & 19.181   \\
 ${13 / 2 }	\,^+$      & 6        &  $6^- \ , \ 7^-$  & 9.376 & 10.729 & 12.010 & 13.233 & 14.406 & 15.536 & 16.629 & 17.689 & 18.719 & 19.723  \\
 ${13 / 2 }	\,^-$      & 7        &  $6^+ \ , \ 7^+$  & 9.376 & 10.729 & 12.011 & 13.234 & 14.407 & 15.537 & 16.630 & 17.690 & 18.720 & 19.724  \\
\end{tabular}
\end{ruledtabular}
\end{table*}

Eq.  (\ref{eq:theorem}) is a general property of any theory or model where
 the quarks are the only degrees of freedom,
the quark-antiquark interaction is chiral invariant and linearly confining at large separations, 
and quarks are light.
Notice  Eq. (\ref{eq:theorem}) only holds if the relevant degrees of freedom are the ones of the light quark. In case the angular excitations in the spectrum would correspond to add other degrees of freedom, say more constituent gluons or quarks, in principle we would have $M^\pm \sim j$, different from Eq. (\ref{eq:trajectory}). Since  linear trajectories similar to Eq. (\ref{eq:trajectory}) hold for different hadron spectra, we are confident in our theorem. 

On the other hand, our assumption that the angular momentum is large and resides in the light quark, can be checked independently from the chiral splitting $\left| M^+ - M^- \right|$, directly in the Regge trajectories.
We now apply the Bohr-Sommerferld quantization rule,  $ k r  \to j$. Then, using the relativistic virial theorem, we find $ \langle  k ^2 \rangle =  \langle  \sigma k  r  \rangle = \langle  \sigma^2 r ^2 \rangle  \to \sigma  j$ and we get the Regge trajectory,
\be
\left( {M^\pm} -m_\text{static} \right)^2 \to 4 \,  \sigma  j \ ,
\label{eq:sigmaalpha}
\ee
where for high $j$ we neglect the intercept $\alpha_0$.
This implies $ \alpha \sim 4 \sigma  $. Thus, measuring the linear leading Regge trajectory in heavy-light mesons should provide an indication of both the degrees of freedom of the system and of the string tension. 
Comparing with Eq. (\ref{eq:trajectory}), this reduces the number of independent parameters, and we get
\be
\left| M^+ - M^- \right| \to    c \ m \left( { 1 \over 2 \sqrt{ \alpha }} \sqrt j \right)  \ .
\label{eq:theorem2}
\ee

Presently, the different approaches to compute the quark dynamical mass 
obtain very diverse quark masses, illustrated in Fig. \ref{fig:massgeneration},
published in Ref. \cite{Bicudo:2009cr}.
Eqs. (\ref{eq:theorem}) and (\ref{eq:theorem2}) provide a gauge invariant tool to measure the momentum dependence 
of the dynamical quark mass.

\begin{figure}[t!]
\includegraphics[width=.95\columnwidth]{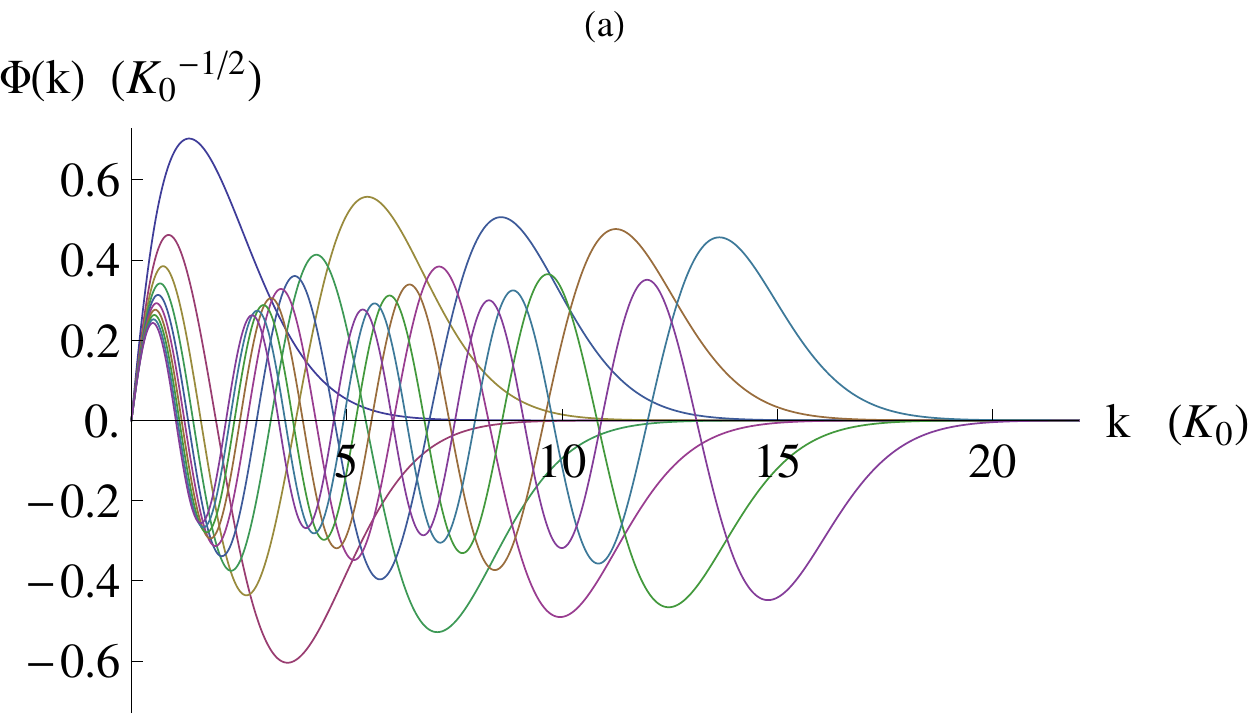}
\hspace{5pt}
\includegraphics[width=.95\columnwidth]{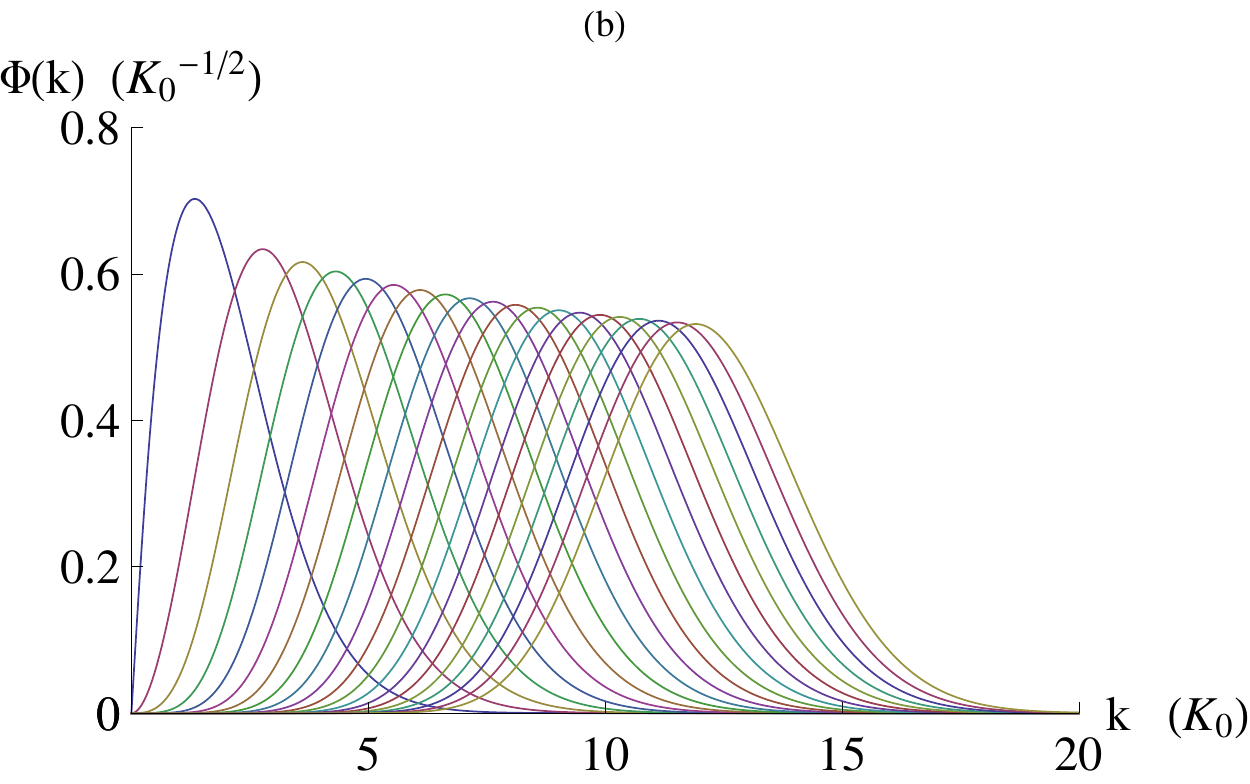}
\caption{
\label{fig:modelwavefu}
(Color online). The quadratic model wavefunctions as a function of momentum in units of $K_0$  (a) for the first 10 radial excitations and (b) for the first 20 angular excitations of the groundstate $j^p=1 / 2 \, ^+$ up to $ 39 / 2 \, ^-$. The radial excited wavefunctions in (b) have a complicated momentum distribution with many peaks and nodes. The angular excited wavefunctions (a) have a well defined momentum distribution with a single peak, and the average momentum clearly increases monotonously with $j$.
}
\end{figure}

\section{$\chi$RS in chiral invariant quark models for antistatic-light mesons  \label{SEC003}}

\subsection{Chiral quark model with a quadratic potential \label{subsec:quadratic}}

To illustrate the mechanism and the quantum numbers for chiral symmetry restoration in the spectrum with a practical example, we utilize the model of  Ref.   \cite{Bicudo:1998bz}.  This model is derived directly form QCD in the local coordinate gauge, also known as Balitski gauge, but it is only approximate due to the truncation of the Dyson-Schwinger equations. The resulting potential is quadratic and not linear as we would expect from confinement in QCD, and in that sense this framework can be regarded as a toy model. This model was first solved in Ref. \cite{Amer:1983qa,LeYaouanc:1983huv,LeYaouanc:1983it}, and the vacuum structure \cite{Bicudo:1989sh}, bound state \cite{LeYaouanc:1984ntu,Bicudo:1989si} and coupled channel equations \cite{Bicudo:1989sj} have also been solved. 

The static-light spectrum of   Ref.   \cite{Bicudo:1998bz} already exhibited $\chi$RS. Here we solve the numerical equations again, in order to improve the numerical precision and perform tests of our $\chi$RS theorem. 

We apply Dyson-Schwinger techniques in Minkowski space, convenient to compute the excited hadronic spectrum. This approach to antistatic-light mesons, clarifies the interplay of the simple quantum operators of the light quark, i e $\mbf l$, $\mbf s$ and $\mbf j$. Ref.   \cite{Bicudo:1998bz} arrives at a Hamiltonian for the light quark, with the quadratic confining potential depending on a single constant $K_0$ related to the Wilson loop and the flux tubes measured in lattice QCD. 
The resulting Salpeter equation for the bound state equation, derived in momentum space, in the chiral limit of a vanishing current quark mass, reads
\begin{eqnarray}
\label{eq:bs}
& &\left\{ E(k) + 2 K_0^3\left(  {1\over 2 k^2} \left[1- {m(k) \over \sqrt{k^2 + m^2(k)}}\right]^2 
\right.\right. 
\\ \nonumber
& & 
\
\left.\left. + {2\over k^2}\left[1-{m(k) \over \sqrt{k^2 + m^2(k)}}\right]
\, {\bf S}\cdot{\bf L}  - \Delta \right) \right\}  \Phi({\bf k}) = M \,  \Phi({\bf k}),
\end{eqnarray}
where $E(k)$ is the quark kinetic energy,
\bea
\label{quarkenergy}
E(k) &=& \frac{k^2+ m_\text{current} m(k)}{\sqrt{k^2+m(k)^2}}
\\ \non 
&& -{K_0}^3\frac{2 \left(k^2+m(k)^2\right)+\left(m(k)-p m'(k)\right)^2}{2 \left(k^2+m(k)^2\right)^2} \ .
\eea
We solve the Salpeter equation with a standard eigenvalue technique, see Table \ref{tab:quadratic} for the resulting spectum of Eq. (\ref{eq:bs}).

Notice the quark mass $m(k)$ is generated consistently with the truncated Dyson-Schwinger equations, to stabilize the physical vacuum\cite{Bicudo:1998bz},
\bea
 m''(k) &=& {2 \over {K_0}^3}
  \sqrt{k^2+m(k)^2} \left[ m(k) - m_\text{current}   \right]
\non \\
&& -2 {m(k) \over k}{ k+2 m(k) m'(k)-k \,  m'(k)^2 \over  k^2+m(k)^2 } \ .
\label{massgap}
\eea 
Hence this simple approach produces a propagator similar to the one in Eq. (\ref{propag}), where the constituent mass $m(k)$ leads to the spontaneous breaking of chiral symmetry, producing a difference between the masses of mesons with opposite parity.

Nevertheless, as anticipated in Section I, chiral symmetry is restored high in the spectrum. This is clear in the spectrum of Table \ref{tab:quadratic}. Notice restoration occurs for large angular momentum quantum number $j$,
\be
\lim_{j\to + \infty} | M^{P=+} - M^{P=-}| = 0 \ .
\label{limlargej}
\ee 
On the other hand, the chiral splitting apparently remains finite with increasing radial excitation quantum number $n$.

The mechanism for $\chi$RS is the following. 
The Hamiltonian includes an angular repulsive barrier and a  spin-orbit interaction. In the limit of large momentum, where $m(k) /k \to 0$, and working with spherical harmonics to decouple the radial momentum $k$ from the angular coordinates,  the spin and angular momentum terms in the Hamiltonian of the Salpeter Eq. (\ref{eq:bs}) are,
\bea
H &=& E(k) +  2\, K_0^3 \left( { 1 \over 2 \, k^2} + {2\over k^2} \,  {\bf s}\cdot{\bf l}  - \Delta \right)
 \non \\  &=&   E(k) + 2\, K_0^3 \left( { 1 \over 2 \, k^2} +
{2\over k^2} \,  {\bf s}\cdot{\bf l}  +  {1\over k^2} \, {\bf l}\cdot{\bf l} - {d^2 \over d k^2} \right)
\non \\ 
&=& k +  2\, K_0^3 \left(- {3 \over  2\,  k^2}+ {1\over k^2} \, {\bf j }\cdot{\bf j} - {d^2 \over d k^2} \right) \ .
\label{resto}
\eea
The final equation is uni-dimensional, in the radial momentum coordinate only. 
Due to the mass independent part of the spin-orbit potential, the Hamiltonian now depends only on $\mbf j \cdot \mbf j =j(j+1)$ and not anymore on  $\mbf l \cdot \mbf l =l(l+1)$ as a non-relativistic quark model would. This implies a degeneracy between parity + and parity - mesons, and thus we obtain $\chi$RS.

Moreover, when the limit $m(k) /k \to 0$ is not exact, a finite splitting  $| M^{+} - M^{-}|$ remains in the spectrum, due to the remaining spin-orbit correction,
\be
-2 \,K_0^3  {2\over k^2}{ {m(k) \over \sqrt{k^2 + m^2(k)}}} \, {\bf s}\cdot{\bf l} \to -{4 \, K_0^3 \over k^2}   { m(k) \over k } 
\, {\bf s}\cdot{\bf l}
\label{eq:m/k}
\ee
and we find that, the breaking of the $\chi$RS is due to a spin-orbit potential, as derived in our theorem, confirming the $ {\bf s}\cdot{\bf l}$ is the only possible tensor potential depending on the angular momentum $\mbf l$ and spin $\mbf s$ of the light quark. Moreover the breaking of the $\chi$RS is proportional as expected to  $m(k) / k $ and to a term with the scale of the interaction, $K_0^3 / k^2$. In the quadratic model, this multiplicative term replaces the term $\sigma / k$ of our theorem.

\begin{table}[t!]
\caption{Meson mass $M$ in  units of the potential parameter $\sqrt \sigma$ as obtained by Ref.   \cite{Sazonov:2014qla}.}
\label{tab:linear}
\begin{ruledtabular}
\begin{tabular}{|c|c|c|c|c|}
$j^p$ & $n=0$ & $n=1$ & $n=2$ & $n=3$
\\
\hline
 ${1/2}^+$ &	1.84	&	2.87	&	3.71	&	4.42	\\
 ${1/2}^-$ &		2.04	&	3.08	&	3.91	&	4.61	\\
 ${3/2}^+$ &	2.84	&	3.67	&	4.38	&	5.01	\\
 ${3/2}^-$ &		2.82	&	3.63	&	4.32	&	4.94	\\
 ${5/2}^+$ &	3.47	&	4.18	&	4.80		&	5.37	\\
 ${5/2}^-$ &		3.48	&	4.19	&	4.82	&	5.39	\\
 ${7/2}^+$ &	4.02	&	4.65	&	5.23	&	5.76	\\
 ${7/2}^-$ &		4.02	&	4.65	&	5.22	&	5.75	\\
 ${9/2}^+$ &	4.49	&	5.07	&	5.60		&	6.09	\\
 ${9/2}^-$ &		4.49	&	5.07	&	5.60		&	6.10		\\
\end{tabular}
\end{ruledtabular}
\end{table}


\subsection{Effective chiral quark model with a linear potential \label{subsec:linear}}

We now illustrate the mechanism and the quantum numbers for chiral symmetry restoration in the spectrum with the effective model introduced in Ref.  \cite{Adler:1984ri} as the first confining and chiral invariant quark model for spontaneous chiral symmetry breaking. This model is similar to the quadratic model of Subsection \ref{subsec:quadratic} but is more effective since the confining potential is linear. 

Notice the dynamically generated running quark mass $m(k$ follows a power law for large momentum $m(k) \propto m^{-4}$, with an excellent Pad\'e fit  \cite{Bicudo:2010qp} aready provided with 3 parameters only,
\be
m(k) = {\sqrt \sigma  \over 6.24900   + 26.7910 \, \sigma ^{1} k ^2 + 17.5059\,  \sigma ^{-2} k^4} \ .
\label{eq:lin4}
\ee

The linear confining potential is observed in static-static potentials in Lattice QCD, computed with Wilson loops. 
Moreover the linear confining potential leads to  linear Regge trajectories, in agreement with experiment. 
The boundstate equation for mesons solved in Refs. \cite{Bicudo:1995kq,LlanesEstrada:1999uh}. 
Recently, the static-light spectrum was computed in Ref. \cite{Sazonov:2014qla}. 

However, with the linear potential, the equations are integral in momentum space, and very large terms cancel numerically, thus an excellent numerical precision is more difficult to achieve. The results of Ref.   \cite{Sazonov:2014qla} are displayed in Table \ref{tab:linear}, and we assume the numerical precision is half of the last digit.

\begin{figure}[t!]
\includegraphics[width=.95\columnwidth]{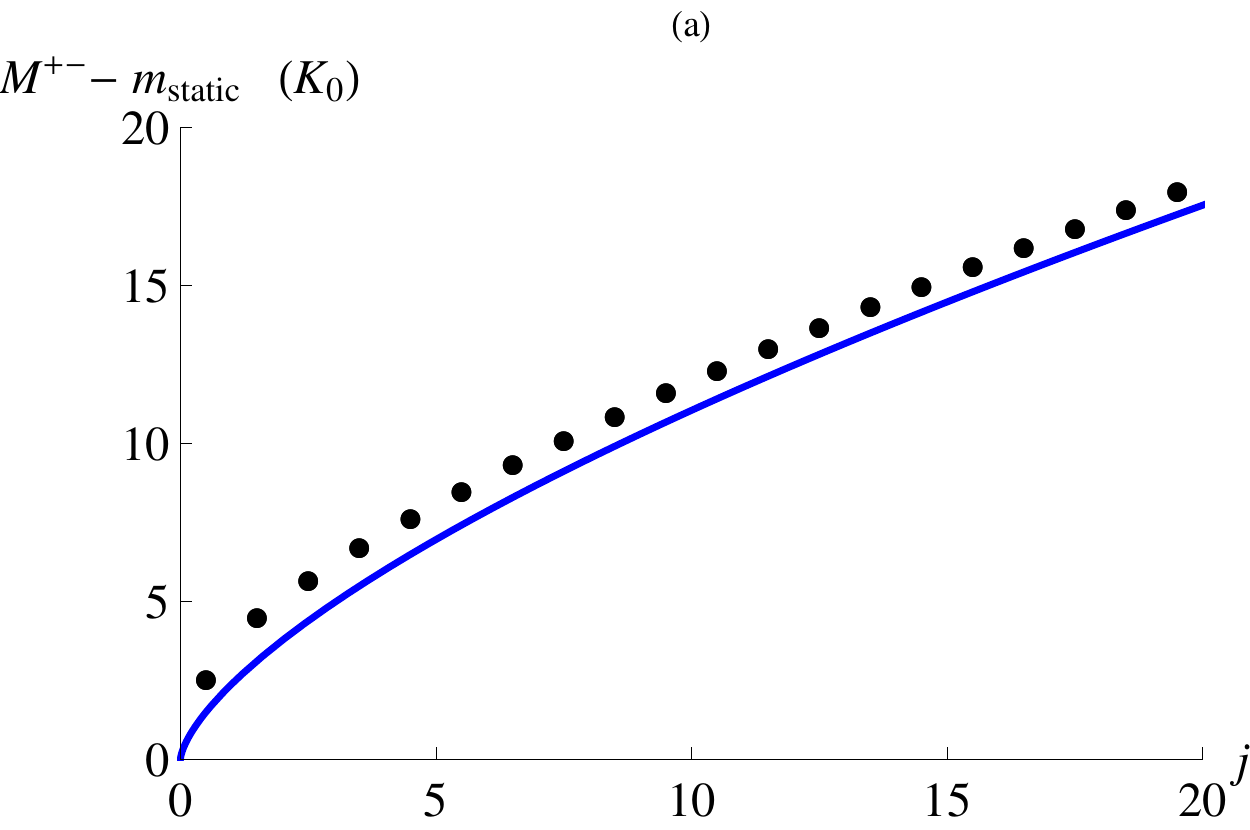}
\hspace{5pt}
\includegraphics[width=.95\columnwidth]{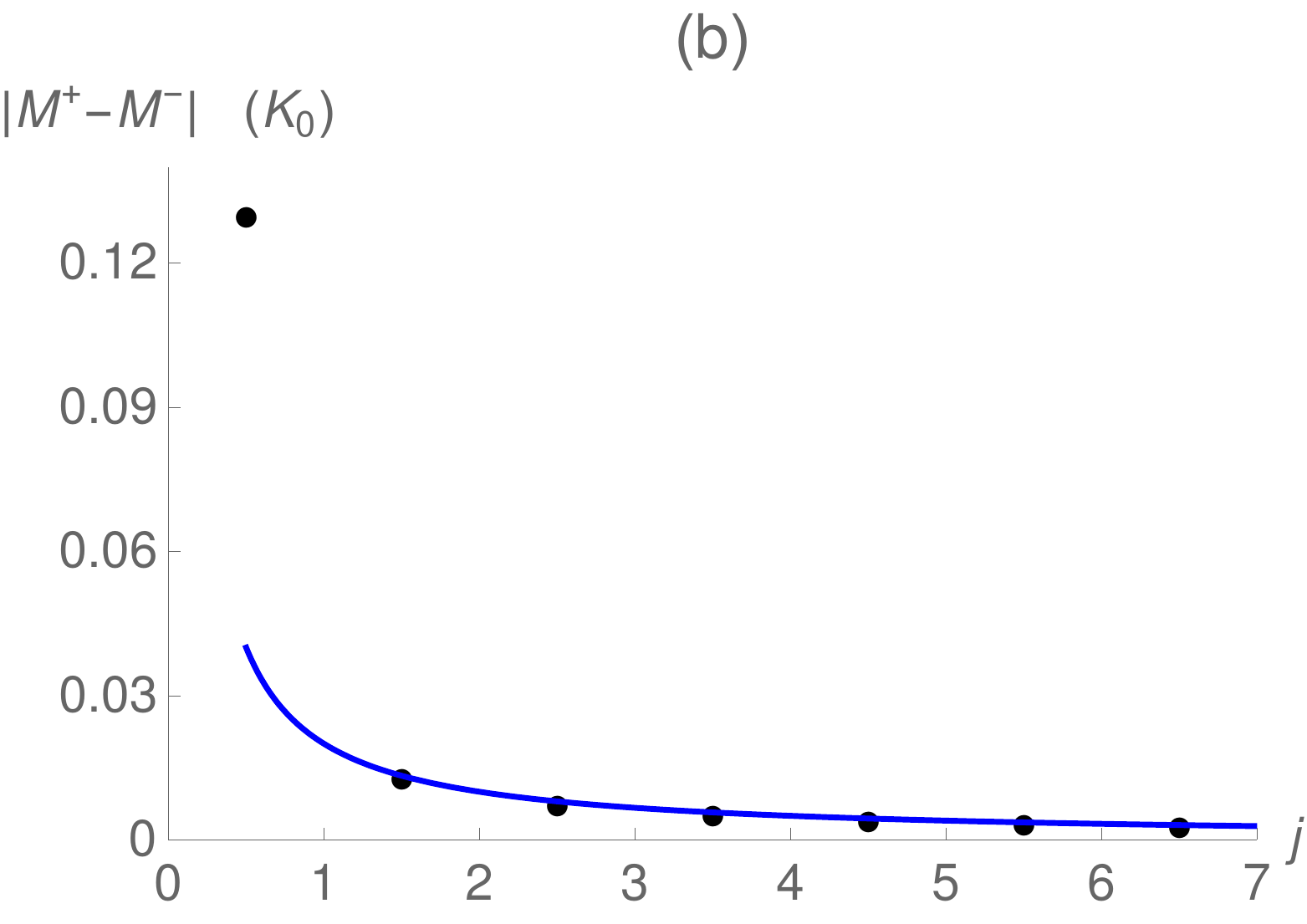}
\caption{
\label{fig:modelM+-2}(Color online). (a) The quadratic model energy static-light energy $\left( {M^\pm} -m_\text{static}  \right)$ and (b) chiral mass splittings $\left| M^+ - M^- \right|$ in units of $K_0$, for angular excited wavefunctions,  as a function of $j$. In solid lines we draw the prediction of our chiral theorem, adapted to the quadratic model. Since the quadratic model's $m(k)$ vanishes too fast for a graphically visible result, in (b) we utilize in the Salpeter equation an arbitrary small mass $m(k)=0.02 K_0$, the agreement with the theorem is excellent already for $j> 1/2$. Moreover we checked numerically our theorem when using the quadratic model's $m(k)$.
}
\end{figure}

\begin{figure}[t!]
\includegraphics[width=.95\columnwidth]{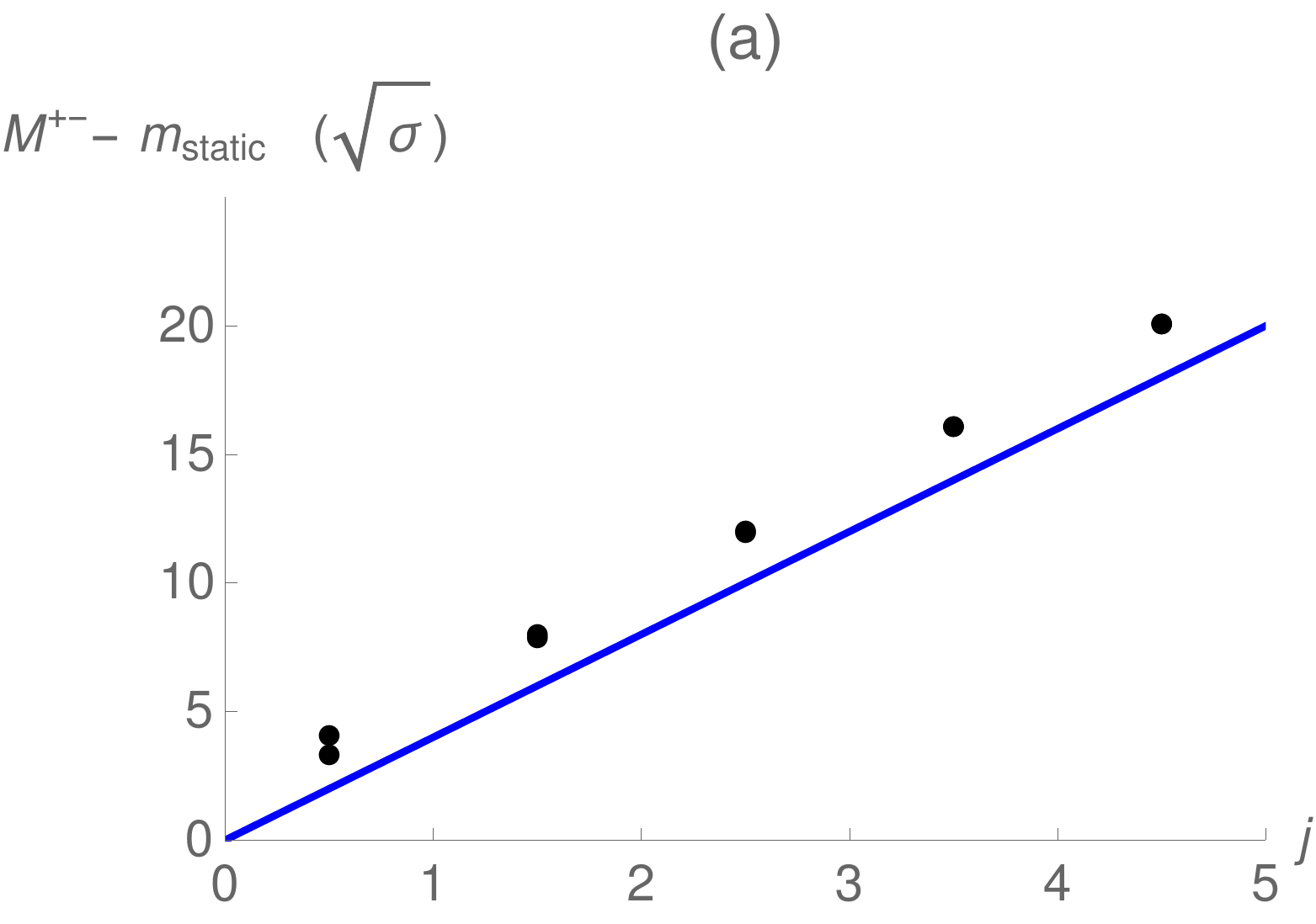}
\hspace{5pt}
\includegraphics[width=.95\columnwidth]{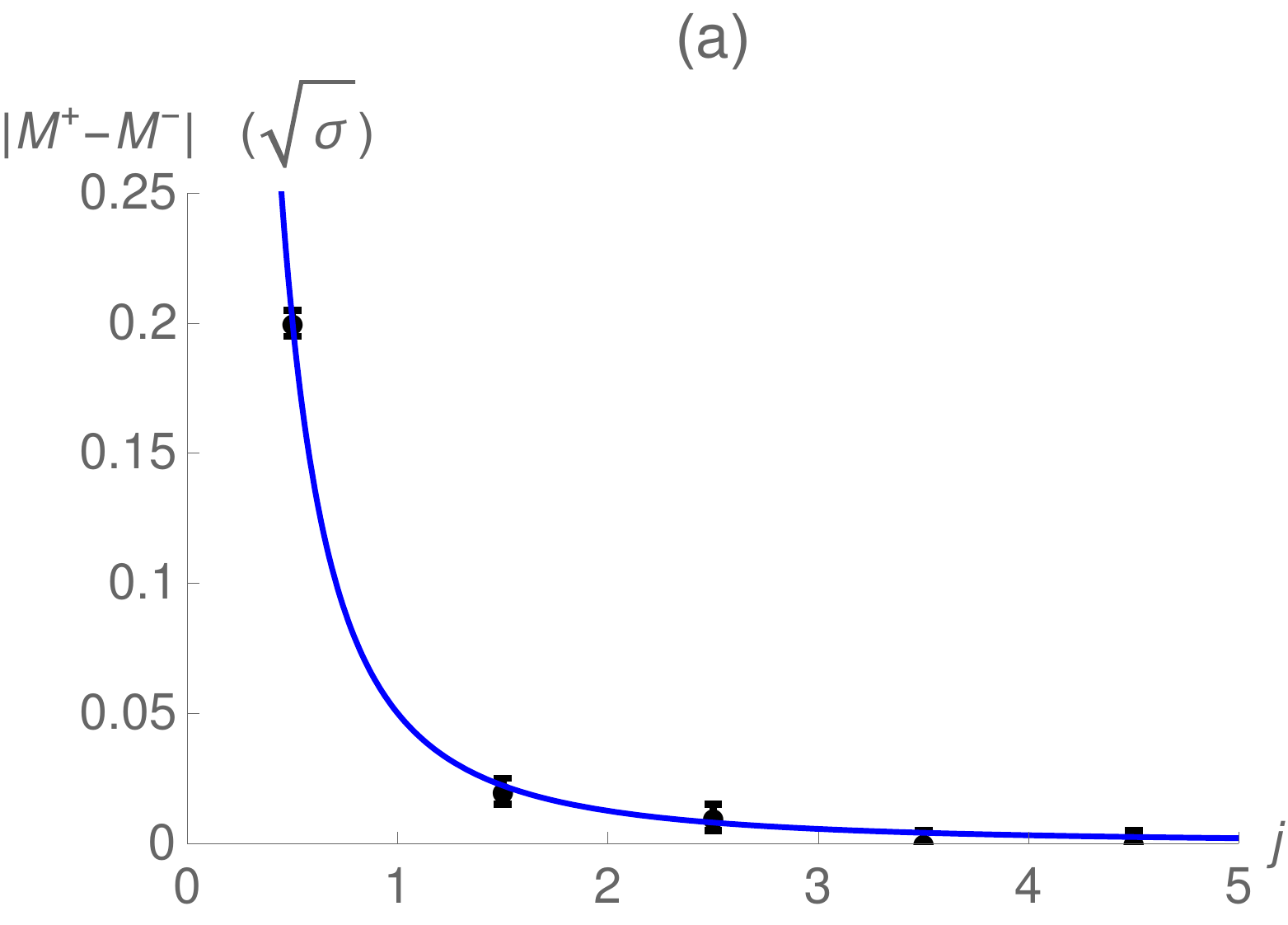}
\caption{
\label{fig:modelwagM+-2}(Color online). (a) The effective linear model energy static-light energy $\left( {M^\pm} -m_\text{static}  \right)$  (b) chiral mass splittings $\left| M^+ - M^- \right|$ in units of $\sigma$, for angular excited wavefunctions,  as a function of $j$. The points are the results of the effective model and in solid lines we draw the prediction of our chiral theorem. 
}
\end{figure}

\subsection{Checking our theorem with the chiral invariant quark models}

\subsubsection{Quadratic model}

We now verify our theorem of Section \ref{SEC002} with the two chiral invariant quark models.

The model used in Subsection \ref{subsec:quadratic}, leading to the boundstate Eqs. (\ref{eq:bs}) , (\ref{resto}) and to the mass Eq. (\ref{massgap}) has a quadratic potential and not linear as we would expect from confinement in QCD. In the large momentum limit, all these equations lead to Airy-like functions, with a decay in $\sim \exp \left(- \text{const} \ k^{3/2} \right)$ for large momenta $k$. For instance, in the limit of large momentum, the mass $m(k)$,
\be
m(k) \to 0.74 \, K_0 \, \exp[ - 0.96 \,  (k /K_0) ^{3/2}] \ ,
\label{eq:massAiry}
\ee  
in the quadratic model decreases extremely fast with $k$ as in an Airy function. 
This is one of the models with the fastest convergence of $m(k) \to 0$ we know of, see Fig. \ref{fig:massgeneration}. 

Nevertheless, feeding the boundstate Eq. (\ref{eq:bs}) with different mass functions $m(k)$ in the model, we get different tests of our theorem of Eqs. (\ref{eq:theorem}) and (\ref{eq:theorem2}) adapted to the quadratic model. Eq. (\ref{eq:paritysplitofm}) is unchanged and the first part of (\ref{eq:averagemomentum}) remains qualitatively correct, but Eqs. (\ref{eq:Hqcd}) and (\ref{eq:trajectory}) must be adapted since this model's confining potential is quadratic. 

Eq. (\ref{eq:Hqcd}) is replaced by Eq. (\ref{eq:m/k}). In what concerns the linear Regge trajectory, it is replaced by another power law relation. The semi-classical Bohr-Sommerfeld quantization of orbits implies
\cite{Bicudo:2007wt}
at large $j$ we have  $ r  \,  k \to j$. Moreover from the light quark Hamiltonian $H \to  k + 2 {K_0}^3 \, r^2$ we find, using the relativistic virial theorem \cite{Lucha:1990mm} we find  both $ \langle k \rangle \sim 2 \langle2  {K_0}^3 \,  r^2 \rangle \sim 4  {K_0}^3 \langle   r \rangle^2$ and 
$  \langle k \rangle \to  { 2 \over 3} M ^{\pm} , \  \langle 2  {K_0}^3 \, r^2  \rangle \to  { 1 \over 3} \left( {M^\pm} -m_\text{static} \right)$. Combining these simple relationships and assuming the probability distributions are well behaved, with a single peak as in Fig. \ref{fig:modelwavefu}, we find the momentum $k$ dependence in $j$,
\be
\langle k / K_0 \rangle  \to  (2 \, j )^{ 2 \over 3} \ ,
\label{eq:modelmom}
\ee
and we find the Regge trajectory for the quadratic model, is no longer linear as in QCD, and is changed to,
\be
{M^\pm} -m_\text{static} \to {3  \over 2^{1 \over 3}} \, j^{ 2 \over 3} \ K_0 \ .
\label{eq:quadregge}
\ee
Moreover, replacing Eq. (\ref{eq:modelmom}) in  Eq. (\ref{eq:m/k}), we finally find, in the quadratic model, the theorem for the mass dependence of the light quark total angular momentum is changed to,
\be
\left| M^+ - M^- \right| \to   { m \left[  \langle k \rangle  (j) \right]  \over j }  \ .
\label{eq:modeltheorem}
\ee
Thus we get an extra power law of $ 1 \over j$ when compared with the result we expect in QCD.

To verify Eqs. (\ref{eq:quadregge}) and (\ref{eq:modeltheorem}) , we plot in Fig. \ref{fig:modelM+-2} the static-light energy $ {M^\pm} -m_\text{static} $ and the chiral mass splitting $\left| M^+ - M^- \right|$ as a function of the total angular momentum $j$. 

In Fig. \ref{fig:modelM+-2} (a), we find a good agreement with Eq (\ref{eq:quadregge}). 
A better fit to the light quark energy, including a next to leading order term to fit the intercept, is  ${M^\pm} -m_\text{static} \to {3  \over 2^{1 \over 3}} \, \left( j +1.33 \right)^{ 2 \over 3} \, K_0$, and thus we find an intercept $\alpha_0 \sim  2 \times  3^{1/3} \, K_0$.

With the running mass $m(k)$ of Eq. \ref{eq:massAiry} we checked numerically that our theorem, Eq. (\ref{eq:modeltheorem}), is accurate. Since the mass of Eq. \ref{eq:massAiry} decerases extremely fast,
for a clearer view of our theorem, in Fig. \ref{fig:modelM+-2} (b) we use an arbitrary constant small mass $m(k)= 0.02 K_0$. Indeed in Fig. \ref{fig:modelM+-2}(b) we find $\left| M^+ - M^- \right| \to m /j$ with an excellent agreement with Eq. (\ref{eq:modeltheorem}) . While for the lowest $j=1/2$ the deviation with the theorem is still large, for the next $j= 3/2, \, 5/2 \cdots$ the large $j$ theorem already reproduces the chiral splitting.

\subsubsection{Effective linear model}

The effective model of  Subsection \ref{subsec:linear} includes linear confinement, with static-light spectrum computed in Ref. \cite{Sazonov:2014qla}, and thus it should directly comply with our chiral theorem.
In Fig. \ref{fig:modelwagM+-2} we first check whether the model produces a linear Regge trajectory, and then test the chiral mass splittings.

In Fig. \ref{fig:modelwagM+-2} (a) the linear Regge trajectory is plotted and indeed the slope $\alpha = 4 \sigma$ we used in our theorem is correct. 
A better fit to the light quark Regge, including a next to leading order term to fit the intercept, is  $\left( {M^\pm} -m_\text{static} \right)^2 \to (2.0 + 4  j ) \sigma$, and thus the static-light spectrum  \cite{Sazonov:2014qla} correspond to an intercept $\alpha_0   \sim 2.0 \, \sigma$.

In Fig. \ref{fig:modelwagM+-2} (b) we check our theorem on the chiral splitting. Indeed we get a perfect fit with 
$\left| M^+ - M^- \right| \to 0.05 \, \sqrt \sigma j^{-2}  $, with the correct power law \cite{Bicudo:2010qp} corresponding to $m (k) \propto k^{-4}$ of Eq. (\ref{eq:lin4}).


\section{Analysis of parity doublets in the lattice QCD spectra \label{SEC004}}


\subsection{Lattice Setup }

The lattice QCD setup is quite different from the continuum calculations with quark models. 
While it is easy to set the heavy antiquark mas as infinite,  we notice several light quark masses have already been studied with present state of the art lattice QCD techniques, and the extrapolation to the limit  of $m_\text{current} \to 0$ has been recently achieved  \cite{,Michael:2010aa,Wagner:2011fs} for the static-light spectrum.
We now detail how the lattice QCD techniques impact on the computed static-light meson spectrum.

\begin{table}[t!]
\begin{center}
\caption{$M(j^\mathcal{P}) - M(S)$ in $\textrm{MeV}$ extrapolated to physical light quark masses, same table as in Ref \cite{Jansen:2008si,Michael:2010aa} . 
\label{tab:marcs}}
\begin{ruledtabular}
\begin{tabular}{|c|c|c|c|c|c||c|}
 $P_-$ & $P_+$ & $D_\pm$ & $D_+$ & $F_\pm$ & $S^\ast$ & $\chi^2 / \textrm{d.o.f.}$ 
 \\
\hline
 $406(19)$ & $516(18)$ & $870(27)$ & $930(28)$ & $1196(30)$ & $755(16)$ & $0.95$ \\
\end{tabular}
\end{ruledtabular}
\end{center}
\end{table}

\begin{table}[t!]
\caption{ Antistatic-light meson mass splittings in MeV computed in lattice QCD \cite{Jansen:2008si,Wagner:2011fs} with dynamical quarks. 
\label{tab:dynamical}}
\begin{ruledtabular}
\begin{tabular}{|c| c c c}
$j$ & $M^+-M_{1/2}^+$ & $M^--M_{1/2}^+$& $\ar M^+\! -\! M^- \ar $ (MeV) \\
\hline
1/2 & 406 (19) & 0 & 406 (19) \\
3/2 &  516(18) & 870(27) & 354 (32) \\
5/2 & 1196(30) &  930(28) & 266 (41) \\
\end{tabular}
\end{ruledtabular}
\end{table}

The static-light spectrum is computed with the correlators illustrated in \ref{fig:04}. The anti quark static source is placed in the lattice with a Wilson temporal line $L$ (thick straight arrow), the operators corresponding to the light quark wavefunction $\Gamma$ are represented by grey disks, and the light valence quark propagator $S$ by a thin arrowed line. Note the lattice QCD propagator is defined in Euclidean position space, whereas the Dyson-Schwinger propagator is defined in momentum space. Moreover, possible sea quark loops are present as well (thin dashed loops), any possible gluon excitations are also included (not represented in the figure), and lattice QCD includes all possible contributions to the angular momentum $\mbf j$ illustrated in Fig. \ref{fig:01}.

The light quark operator wavefunction $\Gamma$ is an appropriate
combination of spherical harmonics and gamma matrices coupling angular momentum and quark
spin to yield well defined total angular momentum $j$
and parity $p$ for the light quark \cite{Jansen:2008si,Michael:2010aa,Wagner:2011fs,Foley:2007ui}.

The space lattice has usually a cubic symmetry, smaller than the rotation group symmetry $SO(3)$. This limits the number of angular excitations which are straightforward to identify. Whereas in the continuum the spherical harmonics may have any angular momentum, in the continuum only rotations multiple of $\pi / 2$ maintain the lattice. In practice, only the groundstate and 2 angular excitations are straightforward to identify. For more excitations, it is necessary to compare degenerate boundstates belonging to different representations, extrapolated to the continuum limit. Due to these difficulties, presently only the light quark angular momenta $j= 1/2, \, 3/2, \, 5/2$ are available in lattice QCD static-light mesons, see Tables Tables \ref{tab:marcs} and \ref{tab:dublins}.  

The energies in the static-light spectrum are computed with effective mass plateaux. The energy splittings in the spectrum are well determined in lattice QCD. However, when computing the static quark potentials with Wilson lines the energy of the groundstate is in general not precisely determined. For instance this occurs in the static quark - static antiquark potential, as a function of their spacial distance $r$ computed with the Wilson loop, where the shape of the potential is stable except for a constant energy shift depending on the lattice spacing.  In particular the Wilson loop is identical to 1 when $r=0$, whereas the potential should be divergent due to the Coulomb term. Thus in Tables \ref{tab:marcs} and \ref{tab:dublins}, the energies are all expressed relatively to the groundstate energy.

\begin{table}[t!]
\begin{center}
\caption{Same Table as in Ref.  \cite{Foley:2007ui}.
The splitting between the ground state in the $G_{1u}$ (S-wave)
  irrep and the other states determined in this analysis by fits to
  the full optimised correlation matrices. The lattice spacing was
  determined from the spin-averaged 1P-1S splitting in charmonium, on
  the same ensemble.  
\label{tab:dublins}}
\begin{tabular}{cccc}
\hline
 channel & O(3) & wave & $\triangle E$ (GeV) \\
\hline
$G_{1u}$, 1st excitation & 1/2 + & S & 0.504(8) \\
$G_{1u}$, 2nd excitation & 1/2 + & S &  0.82(2) \\
\hline
$G_{1g}$, ground state & 1/2 -,  \, 7/2 -  & P & 0.371(6)\\
$G_{1g}$, 1st excitation & 1/2 -,  \, 7/2 - & P &0.76(3) \\
\hline
$H_{g}$, ground state & 3/2 -,  \, 5/2 -,  \, 7/2 - & P & 0.405(6)\\
$H_{g}$, 1st excitation & 3/2 -,  \, 5/2 -,  \, 7/2 -  & P & 0.84(2) \\
\hline
$H_u$, ground state & 3/2 +,  \, 5/2 +& D & 0.706(2)\\
$H_{u}$, 1st excitation &  3/2 +, \,  5/2 + & D & 1.02(2) \\
\hline 
$G_{2u}$, ground state &  5/2 + & D & 0.65(3) \\
$G_{2u}$, 1st excitation &  5/2 + & D & 1.02(6) \\
\hline
$G_{2g}$, ground state & 5/2 -,  \, 7/2 - & F & - \\
\hline
\end{tabular}
\end{center}
\end{table}

\begin{table}[t!!]
\caption{Antistatic-light meson mass splittings in MeV computed in Ref. \cite{Foley:2007ui}, with quenched lattice calculation. 
\label{tab:quenched}}
\begin{ruledtabular}
\begin{tabular}{|c| c c c}
$j$ & $M^+-M_{1/2}^+$ & $M^--M_{1/2}^+$ & $\ar M^+\! -\! M^- \ar $ (MeV) \\
\hline
1/2 & 0.371(6) & 0 &  0.371(6) \\
3/2 &  0.405(6) &  0.706(2) & 0.301(6) \\
1/2* & 0.76(3) & 0.504(8) &  0.26(3) \\
3/2* &  0.84(2) & 1.02(2) & 0.18(3) \\
\end{tabular}
\end{ruledtabular}
\end{table}


\subsection{Lattice QCD results with dynamical quarks}

Ref. \cite{Jansen:2008si,Michael:2010aa} computes the static-light spectrum using dynamical $N_f = 2$ flavour gauge configurations produced by the European Twisted Mass Collaboration (ETMC). Three different values of the lattice spacing $a =$ 0.051 fm, 0.064 fm, 0.080 fm are used together with various values of the pion mass in the range 280 MeV $\leq m_\pi \leq$ 640 MeV. All lattice volumes are big enough to fulfill $ m_\pi  L > 3.2$. The extrapolation to the continuum limit if performed with the different lattice spacings. 

 Importantly to our study, the two parity $p = \pm 1$ groundstates with light quark $j = 1/2 , 3/2 , 5/2$ are clearly determined, computed and extrapolated, not only to the physical current quark mass, but also to the chiral limit of $m_\text{current} \to 0$.

\begin{figure*}[t!]
\includegraphics[width=.95\columnwidth]{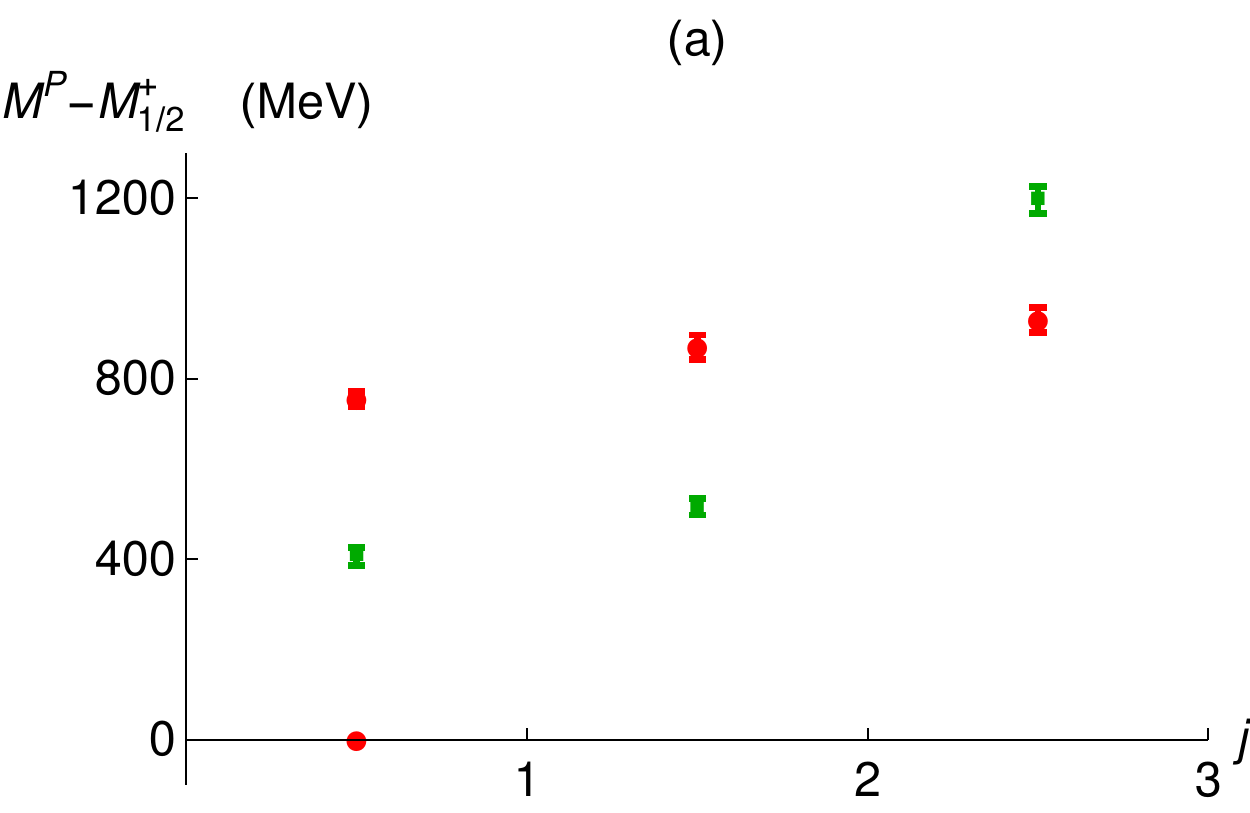}
\hspace{5pt}
\includegraphics[width=.95\columnwidth]{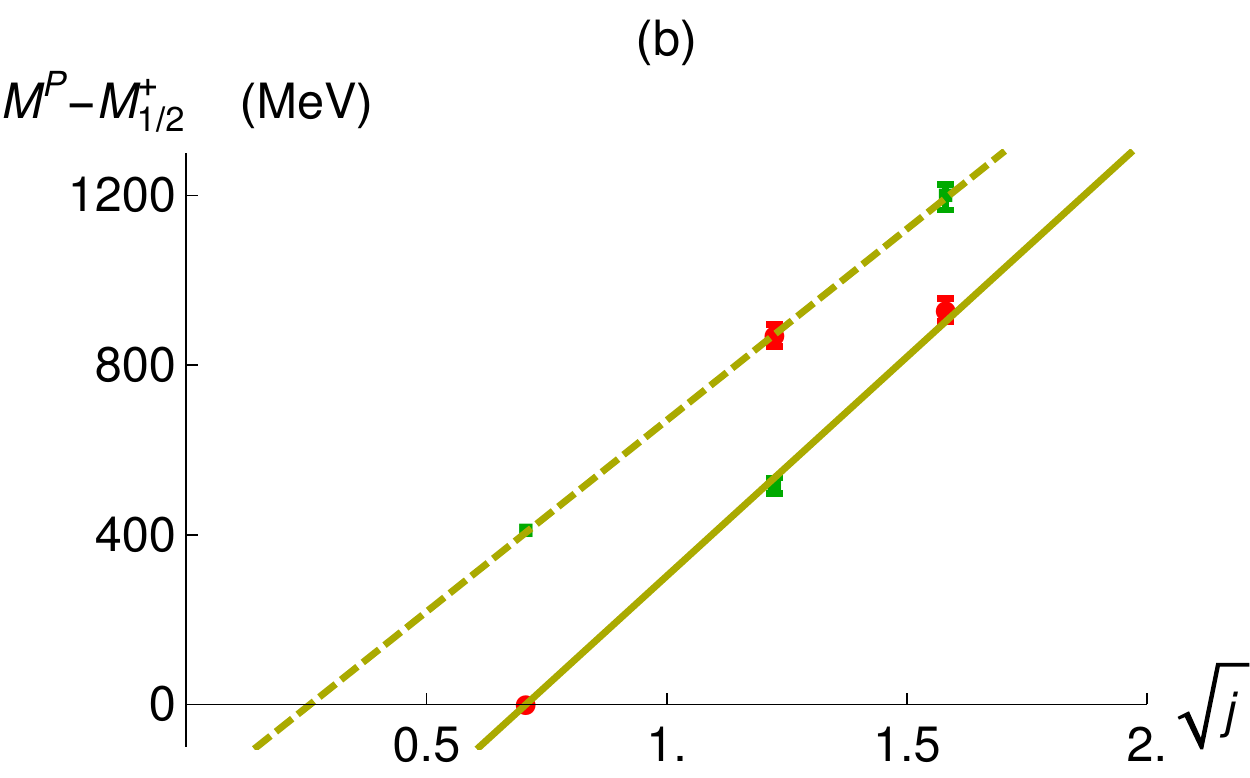}
\\ \vspace{20pt}
\includegraphics[width=.95\columnwidth]{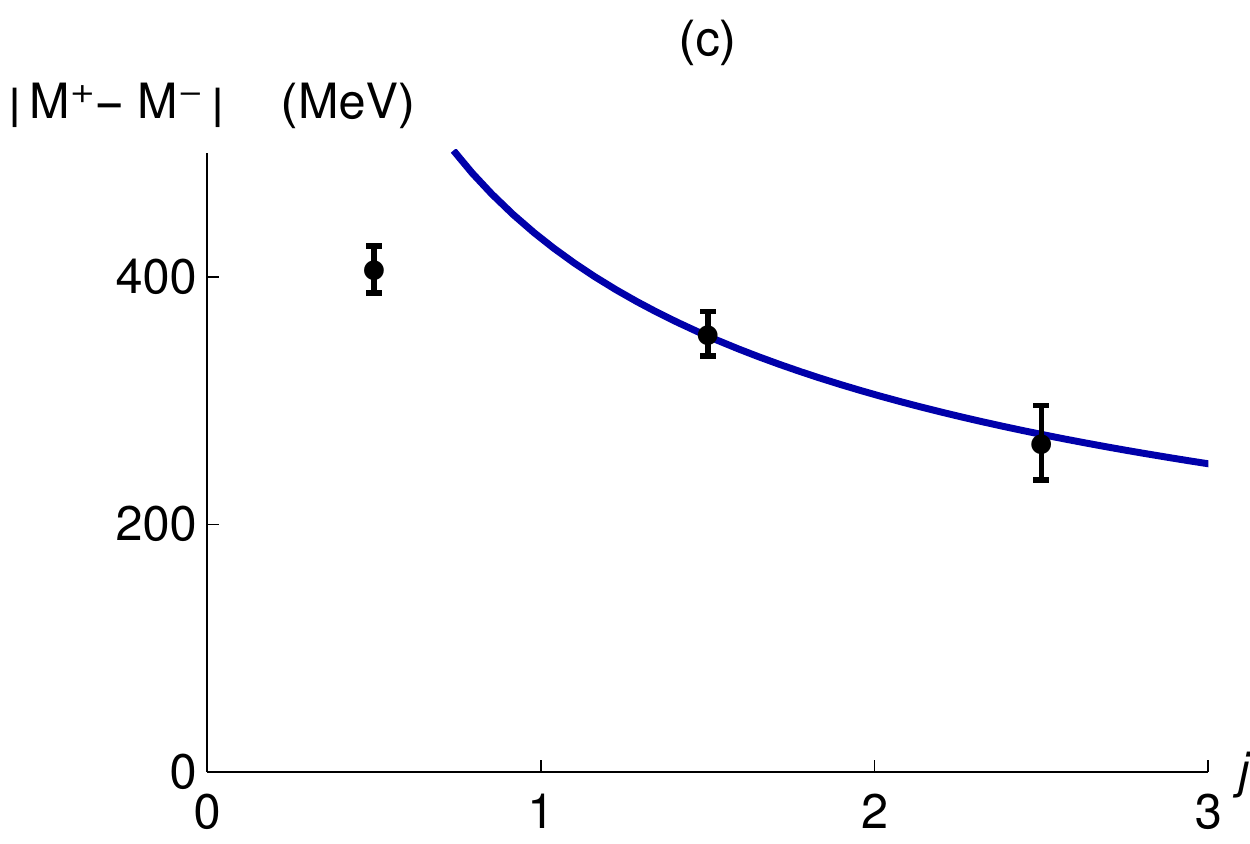}
\hspace{5pt}
\includegraphics[width=.95\columnwidth]{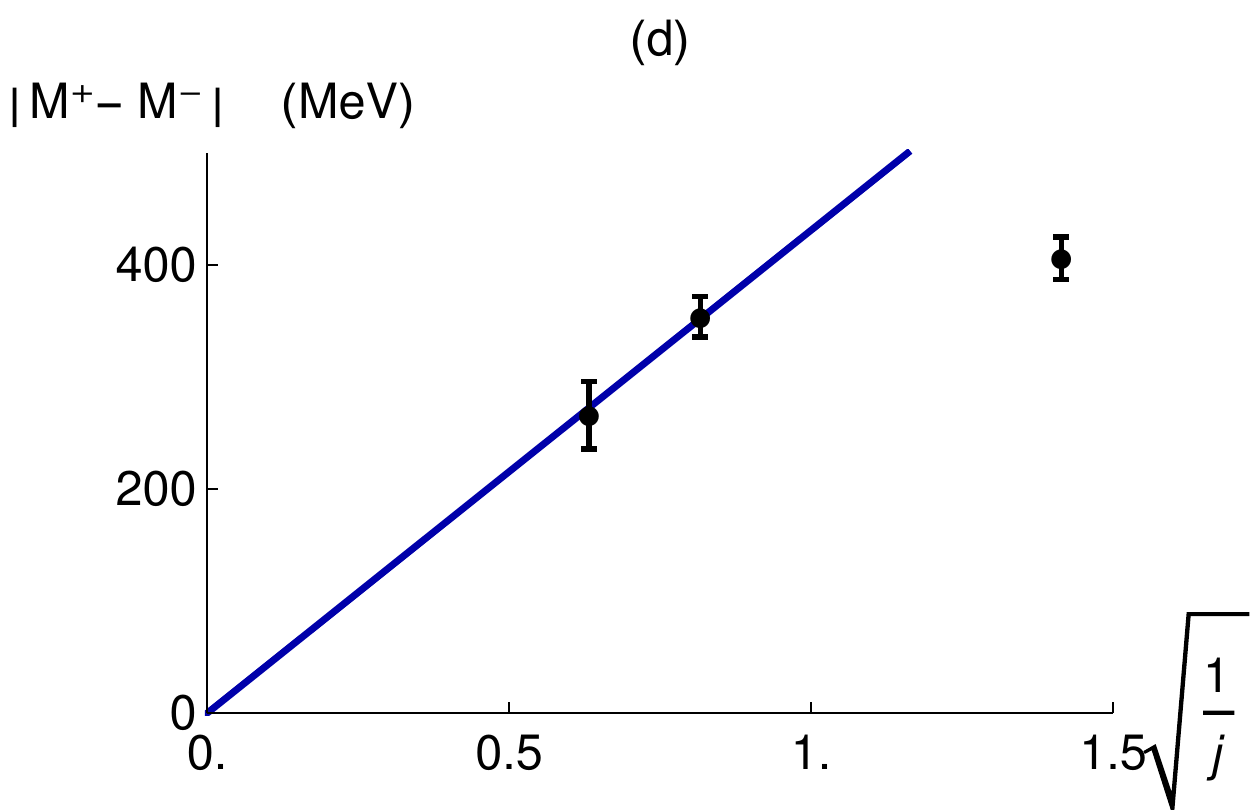}
\caption{\label{fig:latticefit_dynamical}
(Color online) Lattice static-light spectrum computed with dynamical QCD \cite{Jansen:2008si,Michael:2010aa}, and fits.
\\
(a) Spectrum $M^p-M_{1 \over 2}^+$ as a function of $j$, red circles correspond to $p=-$ and green squares correspond to $p=+$.
\\
(b) Regge trajectories $M^p-M_{1 \over 2}^+$ as a function of $\sqrt j$, solid fit for natural $p= j -{1 \over 2}$ and dashed fit for odd $p= j +{1 \over 2}$.
\\
(c) Chiral mass splitting $M^+ - M^-$ as a function of $ j$ and fit $ \propto \sqrt{ 1 \over j}$ of last two points , \
(d) as a function of $\sqrt{ 1 \over j}$. 
}
\end{figure*}

In Tables \ref{tab:marcs} and \ref{tab:dynamical}, and Fig. \ref{fig:latticefit_dynamical} we review the results of Ref. \cite{Jansen:2008si,Michael:2010aa}. All masses in Tables \ref{tab:marcs} and \ref{tab:dynamical} are computed respective to the groundstate, $S$ and s-wave with $J^p={1 \over 2}^+$. The next considered states are,
\bi
\ie
$ S^*$, the radial excitation of the groundstate, an s-wave with $j^p={1 \over 2}^+$, $j=l-{1 \over 2}$,
\ie
$ P_-$, a p-wave with $j^p={1 \over 2}^-$, $j=l-{1 \over 2}$,
\ie
$ P_+$, a p-wave with $j^p={3 \over 2}^-$, $j=l+{1 \over 2}$,
\ie
$ D_\pm$, a d-wave with $j^p={3 \over 2}^+$, $j=l-{1 \over 2}$,
\ie
$ D_+$, a d-wave with $j^p={5 \over 2}^+$, $j=l+{1 \over 2}$,
\ie
$ F_\pm$, a f-wave with $j^p={5 \over 2}^-$, $j=l-{1 \over 2}$,
\ei

In Fig. \ref{fig:latticefit_dynamical} we plot the masses of the static-light mesons in Tables \ref{tab:marcs} and \ref{tab:dynamical}. Moreover we fit the masses in order to study the Regge trajectories and the chiral restoration. 

In Fig.  \ref{fig:latticefit_dynamical}(a) we show the positive and negative parity masses $M^p-M_{1 \over 2}^+$ as a function of $j$. We plot the negative $p=-$ states with red circles and the positive $p=+$ with green squares. The points show a trend compatible both with Regge trajectories and with chiral restoration.

In Fig.  \ref{fig:latticefit_dynamical}(b) we analyse the Regge trajectories. We expect the square of the absolute groundstate masses for each $j$,  $\left(M^p\right)^2$ to be aligned in a linear trajectory with the light quark angular momentum $j$. To obtain the trajectory, 
$ j = \alpha_0 + \alpha M^2$, we would need the absolute value of the masses. However we only have the masses relative to the absolute groundstate $M_{1 \over 2}^+$. Thus to study the trajectory, our best option if to compute $M^p-M_{1 \over 2}^+$ as a function of $\sqrt j$ and fit it with a linear relation. Indeed we find a trajectory with an acceptable linear fit $\chi^2$/ dof=2.02,
\be
M^p-M_{1 \over 2}^+=-730(24) +1033(34) \sqrt j \ \text{MeV} \  .
\ee
The leading trajectory is the one with the lightest masses, with the natural parity $p= j +{1 \over 2}$ corresponding to the angular momenta $j= l +{1 \over 2}$. Moreover we also fit the trajectory with the odd parity, in a perfect linear line, with $\chi^2$ / dof=0.014,
\be
M^p-M_{1 \over 2}^+=-232(5) + 902(4) \sqrt j \ \text{MeV}   \ .
\ee
This suggests the dominant degrees of freedom in these trajectories are the ones of the constituent quark. This is favourable to $\chi$RS.

Figs.  \ref{fig:latticefit_dynamical}(c) and  \ref{fig:latticefit_dynamical}(d) exhibit the power law in $1/k$ of the chiral mass splitting. Using the quadratic model of Section II as a case study, we do not expect the point at $J=1/2$ to allow to study the large $j$ behaviour. Nevertheless, possibly $J=3/2$ and $j=5/2$ are already large enough to observe a quantitative behaviour.   In Fig. \ref{fig:latticefit_dynamical}(c) we plot the antistatic-light meson splittings $\ar M^+ - M^- \ar $, as a function of $j$ and it is clear $\chi$RS occurs monotonously. According to  Eq. (\ref{eq:theorem}), the momentum $k \propto \sqrt j$. Thus in Fig.  \ref{fig:latticefit_dynamical}(d) we plot 
the antistatic-light meson splittings $\ar M^+ - M^- \ar $, as a function of $1 / \sqrt j$. Moreover we test different power law relations for the quark mass $m(k) \propto 1/ k^n$.   
Our fits  suggest $\ar M^+ - M^- \ar \sim c' / \sqrt j$ . We conclude the best fit of the chiral mass splitting , with  ${\chi^2 \over dof}=0.061$,
is the power law
\be
\ar M^+ - M^- \ar = 431 (5) \ j^{-1/2} \ \text{MeV}, \ .
\ee
consistent with a mass depending of momentum with the leading power law at large momentum,
\be
m(k) \propto 1 /k \ ,
\label{eq:powerlaw}
\ee
according to the theorem in Eq. (\ref{eq:theorem}).

\begin{figure*}[t!]
\includegraphics[width=.95\columnwidth]{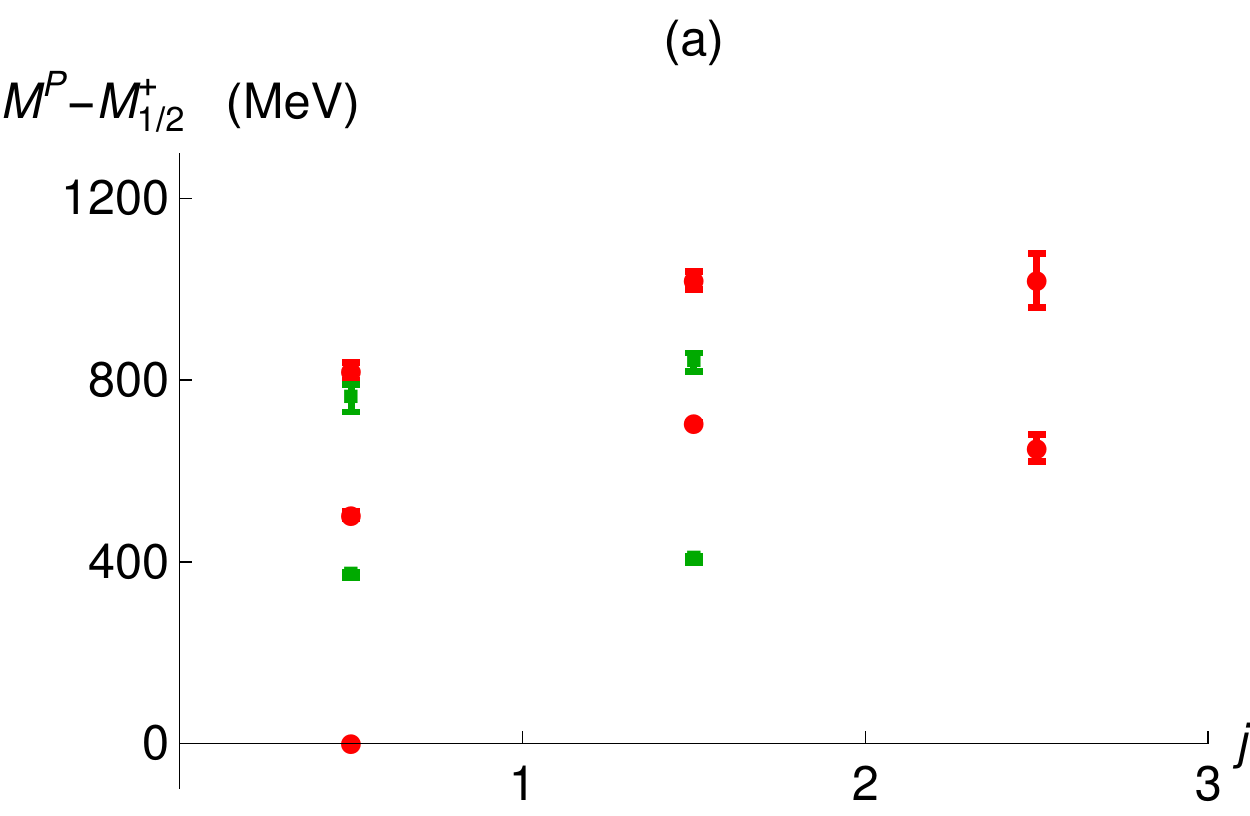}
\hspace{5pt}
\includegraphics[width=.95\columnwidth]{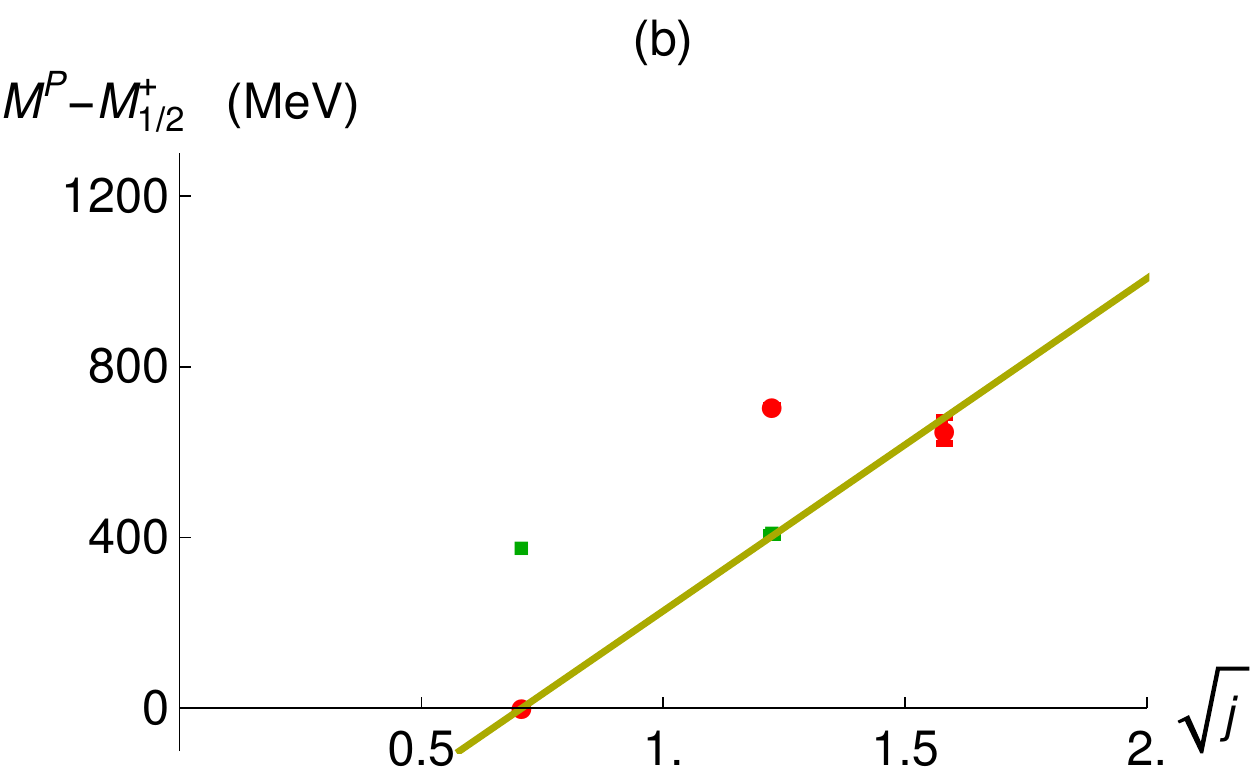}
\\ \vspace{20pt}
\includegraphics[width=.95\columnwidth]{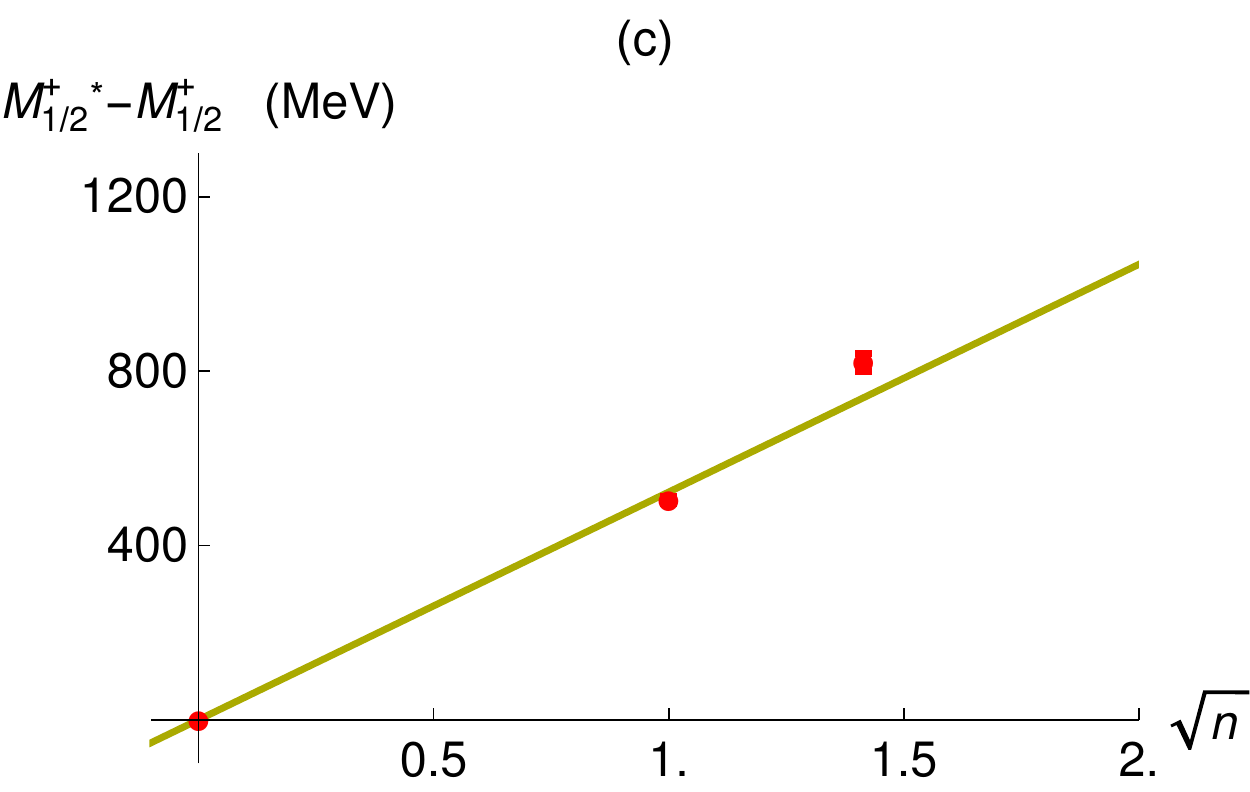}
\hspace{5pt}
\includegraphics[width=.95\columnwidth]{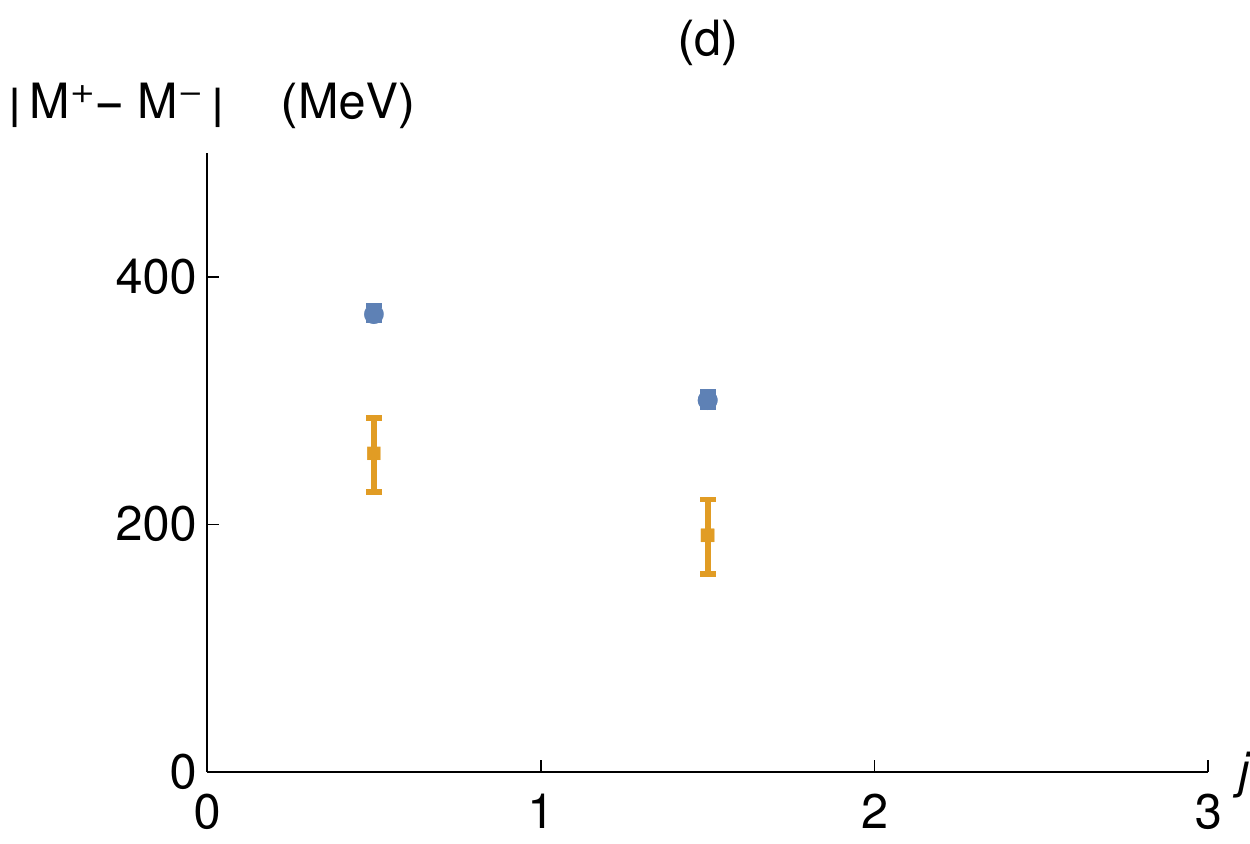}
\caption{\label{fig:latticefit_quenched}
(Color online).  Lattice results with quenched QCD of Ref. \cite{Foley:2007ui}, and fits.
\\
(a) Spectrum $M^p-M_{1 \over 2}^+$ as a function of $j$, red circles correspond to $p=-$ and green squares correspond to $p=+$.
\\
(b) Regge trajectories with  $M^p-M_{1 \over 2}^+$ as a function of $\sqrt j$ and fit of the natural  natural $p= j -{1 \over 2}$ in a solid line.
\\
(c) Regge trajectory in the radial excitations with $M^p-M_{1 \over 2}^+$ as a funtion of $\sqrt n$, and fit in a solid line  .
\\
(d) Chiral splitting $\ar M^+ - M^- \ar$ as a function of $j$ in blue circles, and purple squares for the first excitation splitting. 
}
\end{figure*}


\subsection{Lattice QCD results with quenched quarks}

It is interesting to compare static-light mesons computed in quenched versus dynamical lattice QCD. The differences can be traced to sea quarks. In principle, if quark loops are important for the spectrum, the differences should be noticeable. In Fig. \ref{fig:latticefit_quenched} and Tables \ref{tab:dublins} and  \ref{tab:quenched} we report the results of the quenched calculation of  Ref.  \cite{Foley:2007ui}. 

In this case the continuum limit is not reached, and lattice effects of the octahedral group $O_h$ in the lattice may still be present.
The irreducible representations of the group $SO(3)$ in the continuum, do not coincide with the irreducible representations of the octahedral group in the lattice. Continuum states with a $j_z$ angular momentum continuum quantum number are in general
divided between the lattice irreducible representations. For example, states
with quantum numbers 
$j^{p} = {5  \over 2}^+$
, corresponding to a
6-dimensional representation of $O(3$), appear in both the
$G_{2g}$ and $H_g$ octahedral irreducible representations . The continuum degeneracy is broken by lattice artefacts.
In a numerical study one hopes to determine the 
$5/2+$
energy levels by identifying near-degenerate levels
in the $G_{2g}$ and $H_g$ irreducible representations which converge in the approach
to the continuum limit.

In the notation of Ref.  \cite{Foley:2007ui} , the correspondence between the octahedral irreducible representations and the continuum representations is the following.
\bi
\ie
The s-wave,
which has 
$j^{p}  ={1  \over 2}^+ $, lies in the $G_{1u}$ irrep. 
\ie
The 
${1  \over 2}^- $
p-wave and the 
${3  \over 2}^- $
p-wave appear in the $G_{1g}$ and $H_g$ irreps
respectively. 
\ie
The d-wave multiplets are labeled 
${3  \over 2}^+ $
and 
${5  \over 2}^+ $. 
Both of these appear in the $H_u$ irrep, and the
${5  \over 2}^+$
states also arise in the $G_{2u}$ irrep. 
\ie
The lowest-lying
states in the remaining lattice irrep , the $G_{2g}$ representation,
are expected to have the quantum numbers 
$5 /2-$,
corresponding to f-wave excitations.
\ei
The respective masses are summarized in Table \ref{tab:dublins}. In principle, for the groundstates of each irreducible representation, the corresponding continuum quantum numbers are clear. Possibly the same quantum numbers can be maintained for the first excited state, assuming it is a radial excitation with the same quantum numbers as the groundstate. To identify higher angular momenta, a very careful continuum limit and studies of the degeneracies between $O_h$ irreducible representations would be necessary, and thus we do not pursue it. We arrive at the spectrum depicted in 
 Fig. \ref{fig:latticefit_quenched}(a). The spectrum shows evidence for Regge trajectories in the groundstates for each angular momentum.
 
Again, to study the leading trajectory, our best option if to compute $M^p-M_{1 \over 2}^+$ as a function of $\sqrt j$ and fit it with a linear relation. In Fig. \ref{fig:latticefit_quenched}(b) we plot the trajectories. For the 
leading trajectory, the one with the lightest masses, with the natural parity $p=(-1)^{ j -1 / 2}$ corresponding to the angular momenta $j= l +{1 \over 2}$, there are three points. Indeed we are able to fit them with a linear trajectory, with $\chi^2/dof=1.14 $,
\be
M^p-M_{1 \over 2}^+=-550 (8)  + 778	(12)  \sqrt j \ \text{MeV} \ .
\ee
However, for the odd parity trajectory, we only have two points, insufficient to compute the $\chi^2/dof$ with a linear fit. Nevertheless, this trajectory is essentially parallel to the leading one, and this suggests, as in the dynamical case, both may have a comparable slope.

In the quenched case, it is interesting as well to study the Regge slope for radial excitations, since we have two excitations of the groundstate with $j^p={1 \over 2}^+$. In Fig. \ref{fig:latticefit_quenched}(c) we plot the masses as a function of $\sqrt n$. As in the chiral quark models of Section \ref{SEC003}, we assume $n=0$ for the groundstate and $n=1, \ 2$ for the two excited states. With power laws in $\sqrt n$, the best fit we find is a linear one, though it is not very good fit, with $\chi^2/dof=21.7 $
\be
M^p-M_{1 \over 2}^+=522 (32) \sqrt n \ \text{MeV} \ .
\ee
Although the $\chi^2/dof$ is large, it is nevertheless interesting to notice the radial Regge slope $522 (32)$ is of the same order of the one of the angular Regge slope $550 (8)$. This is an interesting result, possibly pointing to hybrid effects,  e g quantum numbers beyond the valence quark ones \cite{Bicudo:2007wt}, in the radial excited spectrum \cite{Bugg:2004xu}, as observed at the the former Cristal Barrel collaboration at CERN \cite{Aker:1992ny,Anisovich:2000ut,Anisovich:2011sva,Anisovich:2001pn,Anisovich:2002su} for light-light mesons.

Moreover,  in Fig. \ref{fig:latticefit_quenched} (d), where we plot the chiral mass splittings of Table \ref{tab:quenched}, there is a clear evidence of $\chi$RS, between the groundstate trajectories with natural and off parities. We do not attempt to fit quantitatively the amount of restoration, because we only have two points. 
Nevertheless it is also interesting to note there is also evidence,  in Fig. \ref{fig:latticefit_quenched} (d),  for chiral restoration in the first excited trajectories.

Comparing the quenched and dynamical results, there is no strong qualitative difference. The only quantitative difference is  in the slope in the Regge trajectories. This suggests the valence light quark degrees of freedom dominate the angularly excited spectrum of static-light systems.


\section{Conclusion \label{SEC005}}

The static-light mesons constitute an ideal system to test new ideas and theorems in QCD. We specialize the theorem of Ref \cite{Bicudo:2009cr} to the static-light system.

We compute, in a two different chiral invariant quark models, with quadratic and linear confinement, the chiral mass splittings between positive and negative parity partners $\ar M^+\! -\! M^- \ar $ in the antistatic-light spectrum, up to  $j^p= 39 / 2 \, ^\pm$. We utilize the chiral models as benchmarks to test our theorem.

We apply the theorem in order to extract the quark running mass $m(k \propto \sqrt j)$ from the available lattice data.
We find a trend of decreasing $\ar M^+\! -\! M^- \ar $as a function of $j$. This is a signal for $\chi$RS, a very interesting result for the first time identified in a lattice QCD spectrum. 
Moreover the linearity of Regge trajectories, together with the $\chi$RS, suggest the valence/constituent quark degrees of freedom are the dominant ones in the leading trajectories of static-light systems.

 We expect  $\ar M^+\! -\! M^- \ar $ to produce $  \langle  m ( k \propto \sqrt j ) \rangle$. While this cannot be applied to the very small $j = 1 /2$, considering the next two values of $j=3/2$ and $j=5/2$ we find in Lattice QCD an indication $\langle  m ( k ) \rangle \propto 1/\sqrt j$. This is equivalent to a a power law of $m(k) \propto k^{-1}$, explaining how the running quark mass interpolates between the light constituent quark mass of 300 to 400 MeV to the much smaller current quark masses.  Due to the difficulties of the cubic / octahedral symmetry, there are no available static-light lattice QCD masses for  $j > 5/2$.

This momentum dependence of the quark mass is very interesting because it is qualitatively quite different from the one of constituent quark models, of chiral quark models, of the Nambu and Jona-Lasinio model and of Dyson-Schwinger approaches, shown in Fig. \ref{fig:massgeneration}. We expect the present result to lead to further progress of the study of the running mass $m(k)$ in chiral models of QCD. 

Notice $\ar M^+\! -\! M^- \ar $ is gauge invariant, in a sense it provides a gauge invariant definition of the quark running mass. In the future, it would be very interesting to extend the static-light lattice data to higher total angular momenta $j > 5/2$. 

It would be tantalizing for the QCD community if the experimental collaborations, say LHCb at CERN,  could compute the spectrum of more $B$ mesons, so far we only have three confirmed states in the leading trajectories, with $J^P= 1^-$ and $J^P= 2^+$ in the natural parity trajectory and $J^P= 1^+$ in the inverse parity trajectory \cite{Agashe:2014kda}. At least the double of states would be necessary to compare with the lattice QCD spectrum.


\acknowledgements

P.B. is grateful for very important discussions with Marc Wagner on the lattice QCD techniques, symmetries  and results for heavy-light mesons, and acknowledges the support of HIC for FAIR and CFTP (grant FCT UID/FIS/00777/2013), and is thankful for the hospitality of IFT, and to the Physics department of IST for a sabbatical leave while this work was produced.



\end{document}